\UseRawInputEncoding
\documentclass[journal=jpcafh,manuscript=article]{achemso}

\usepackage{longtable}
\usepackage{multirow}
\usepackage{amsmath}
\usepackage{subfigure}
\usepackage[usenames,dvipsnames]{color}
\usepackage{soul}
\usepackage{bm}
\usepackage{rotating}

\usepackage{amssymb}
\usepackage{titlecaps}
\Addlcwords{a an the of on at with for and in into from by without}
\usepackage{soul}

\makeatletter
\newcommand*{\addFileDependency}[1]{
  \typeout{(#1)}
  \@addtofilelist{#1}
  \IfFileExists{#1}{}{\typeout{No file #1.}}
}
\makeatother

\newcommand*{\myexternaldocument}[1]{%
    \externaldocument{#1}%
    \addFileDependency{#1.tex}%
    \addFileDependency{#1.aux}%
}
\usepackage{xr}
\myexternaldocument{si}

\usepackage{multicol} \usepackage{wrapfig} \usepackage{enumitem}
\usepackage{bm} \usepackage{outlines} 
\SectionNumbersOn

\author{Meenu Upadhyay and Marco Pezzella}
\affiliation[University of Basel]{Department of Chemistry, University
  of Basel, Klingelbergstrasse 80 , CH-4056 Basel, Switzerland.}
\author{Markus Meuwly}
\affiliation[University of Basel]{Department of Chemistry, University
  of Basel, Klingelbergstrasse 80 , CH-4056 Basel, Switzerland.}
\email{m.meuwly@unibas.ch}

\title{Genesis of Polyatomic Molecules in Dark Clouds: CO$_2$
  Formation on Cold Amorphous Solid Water}

\begin{document}

\date{\today}

\begin{abstract}
Understanding the formation of molecules under conditions relevant to
interstellar chemistry is fundamental to characterize the chemical
evolution of the universe. Using reactive molecular dynamics
simulations with model-based or high-quality potential energy surfaces
provides a means to specifically and quantitatively probe individual
reaction channels at a molecular level. The formation of CO$_2$ from
collision of CO($^1 \Sigma$) and O($^1$D) is characterized on
amorphous solid water (ASW) under conditions typical in cold molecular
clouds. Recombination takes place on the sub-nanosecond time scale and
internal energy redistribution leads to stabilization of the product
with CO$_2$ remaining adsorbed on the ASW on extended time
scales. Using a high-level, reproducing kernel-based potential energy
surface for CO$_2$, formation into and stabilization of CO$_2$ and COO
is observed.
\end{abstract}

\clearpage

\noindent
The formation and chemical evolution of stars and galaxies is
intimately linked to the presence of molecules.\cite{tielens:2013}
This ``Molecular Universe''\cite{Tielens82p245,tielens:2013} requires
molecules to be generated ``bottom-up'' but also involves them to be
destroyed ``top-down'' by reaction channels that are potentially
specific to the conditions in the interstellar medium. For the
formation of neutrals, grain-surface chemistry involving interstellar
ices inside dense molecular clouds is of particular
importance.\cite{tielens:2013} Reactions on icy surfaces occur at the
diffusion limit and their efficiency depends primarily on the
diffusion of one or several of the reacting species on the surface and
whether or not particular reaction channels involve entrance
barriers. In dense cold clouds, CO is the main carbon
reservoir\cite{tielens:2013} and the second most abundant molecule in
Molecular Clouds with an important role as a molecular tracer for
probing and characterizing the chemical and physical conditions of the
environment.\cite{neininger:1998}\\

\noindent
Observations indicate that the abundances of H$_2$O and CO$_2$ on
grains within clouds are sufficient to be detected in the interstellar
medium.\cite{shenoy:2001,garrod:2011} The chemical precursor for
formation of CO$_2$ is believed to be carbon
monoxide. Thermoluminescence experiments of added CO to photolyzed
N$_2$O in an Argon matrix at 7 K suggested that the O($^3$P)+CO($^1
\Sigma^+$) reaction yields excited CO$_2^*$ which, after emission of a
photon, leads to formation of CO$_2$.\cite{pimentel:1979} Such a
process has also been proposed to occur on interstellar
grains\cite{herbst:2001} and confirmed
experimentally\cite{dulieu:2013} with an estimated entrance barrier of
0.014 eV to 0.103 eV for the process on ASW, compared with a value of
0.3 eV from high-level electronic structure
calculations.\cite{MM.co2:2021} The surrounding water matrix should
facilitate relaxation of the $^3$A$'$ or $^3$A$''$ states of CO$_2$ to
the $^1$A$'$ ground state (correlating with linear $^1 \Sigma_{\rm
  g}^+$). On the other hand, the presence of an entrance barrier for
the O($^3$P)+CO($^1 \Sigma^+$) reaction has one led to consider the
alternative CO+OH pathway for CO$_2$ formation\cite{linnartz:2011}
which was, however, reconsidered to yield the HOCO intermediate in
such environments in more recent experiments.\cite{linnartz:2019}\\

\noindent
For atomic oxygen the possibility for diffusion on and within
amorphous solid water (ASW) was demonstrated down to temperatures of
10 K and 50 K from both, experiments and
simulations.\cite{oxy.diff.minissale:2013,MM.oxy:2014,MM.oxy:2018}
Typical temperatures of dense cold clouds have been estimated at $\sim
15$ K.\cite{bergin:2007} The oxygen mobility found even at such low
temperatures\cite{oxy.diff.minissale:2013} opens up ways to form small
molecules, including O$_2$, O$_3$, or CO$_2$ through atom/atom or
atom/molecule recombination. In this context it is also of interest to
note that excited O($^1$D) state has been reported to be considerably
more mobile than ground state
O($^3$P).\cite{apkarian:1993,apkarian:1999} For molecular oxygen it
was shown that both, ground and excited state O$_2$ can be stabilized
on amorphous solid water.\cite{MM.oxy:2019,MM.o2:2020}\\

\noindent
Direct, barrierless recombination into the ground state of CO$_2$ is
possible from O($^1$D)+CO($^1 \Sigma^+$). Excited O($^1$D) can be
formed from photolysis of H$_2$O\cite{klemm:1975} and the radiative
lifetime is 110 minutes.\cite{garstang:1951} Also, neutral
dissociation of water into H$_2$ and O($^1$D) in the presence of CO
has recently been reported to lead to formation of CO$_2$ in cryogenic
films containing CO and H$_2$O.\cite{schmidt:2019} Here, the reaction
O($^1$D)+CO($^1 \Sigma^+$) $\rightarrow$ CO$_2$($^1 \Sigma_{\rm g}^+$)
on the surface of ASW is investigated at a molecular level. Two
potential energy surfaces (PESs) are used to describe the energetics
of CO+O $\rightarrow$ CO$_2$ formation (see Supporting Information for
the energy functions used). One is a Morse-Morse-Harmonic (MMH)
parametrization based on multi-reference CI calculations for
CO-dissociation, fitted to a Morse functional form together with an
empirical parametrization for the OCO angle $\theta$. The second PES
is a 3-dimensional reproducing kernel Hilbert space (RKHS)
representation of CCSD(T)-F12 calculations which (see Figure
S1 for the quality of the representation) is more
accurate than the MMH PES but also considerably more computationally
expensive to evaluate. Energy conservation for MD simulations using
both PESs is reported in Figure S2.\\

\noindent
The 3-dimensional MMH-PES, fit to MRCI/aug-cc-pVTZ data, is
qualitatively correct.\cite{MM.co2:2021} The equilibrium geometry is a
linear O-C-O configuration with C-O distances of 1.1644 \AA\/ with
bending, symmetric and asymmetric stretching frequencies at 645
cm$^{-1}$, 1226 cm$^{-1}$ and 2450 cm$^{-1}$, respectively, compared
with 667 cm$^{-1}$ for bend, 1333 cm$^{-1}$ for symmetric stretch and
2349 cm$^{-1}$ for asymmetric stretch.\cite{shimanouchi:1978}.\\

\begin{figure}[H]
\centering\includegraphics[scale=0.8]{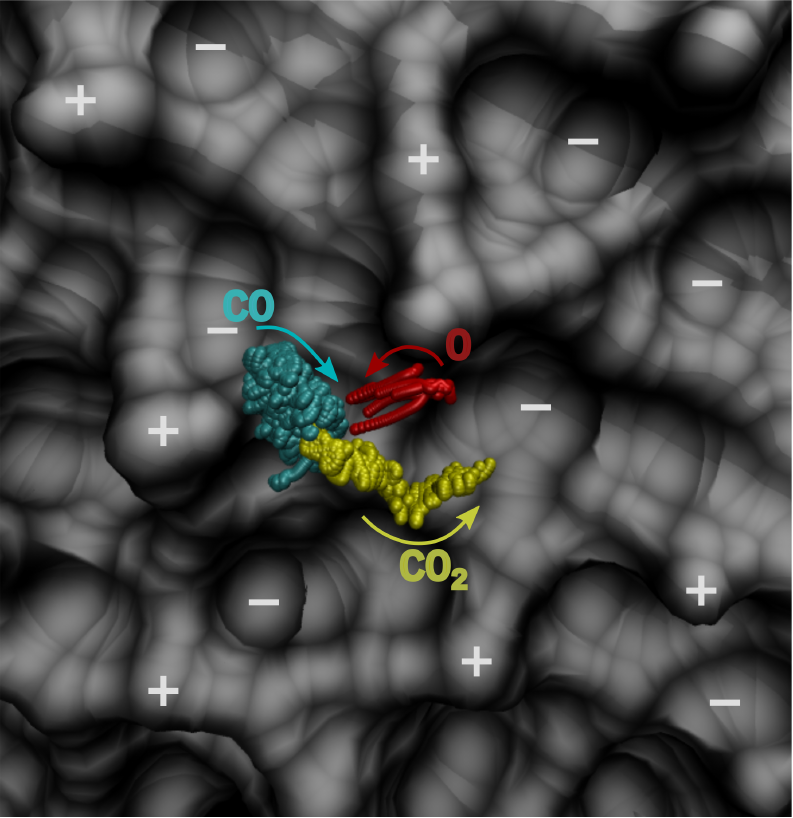}
\caption{Trajectories leading to CO$_2$ recombination. The O-atom
  (red) and CO molecule (cyan) from 8 different trajectories are shown
  to form CO$_2$ (yellow). Before and after recombination all species
  diffuse on the ASW (grey colors) surface. The ``plus'' and ``minus''
  signs indicate protuberances and indentations of the ASW surface.}
\label{fig:struc}
\end{figure}

\noindent
First, recombination simulations were run with the MMH model. This
provides a qualitatively correct description of the CO$_2$ formation
dynamics at reduced computational cost. Typical recombination
trajectories projected onto the ASW surface are shown in Figure
\ref{fig:struc}. Before recombination, CO (cyan) and O (red) diffuse
separately in their respective adsorption sites on the ASW. After
recombination, CO$_2$ (yellow) continues to diffuse on the ASW.\\

\begin{figure}
\centering \includegraphics[scale=0.54]{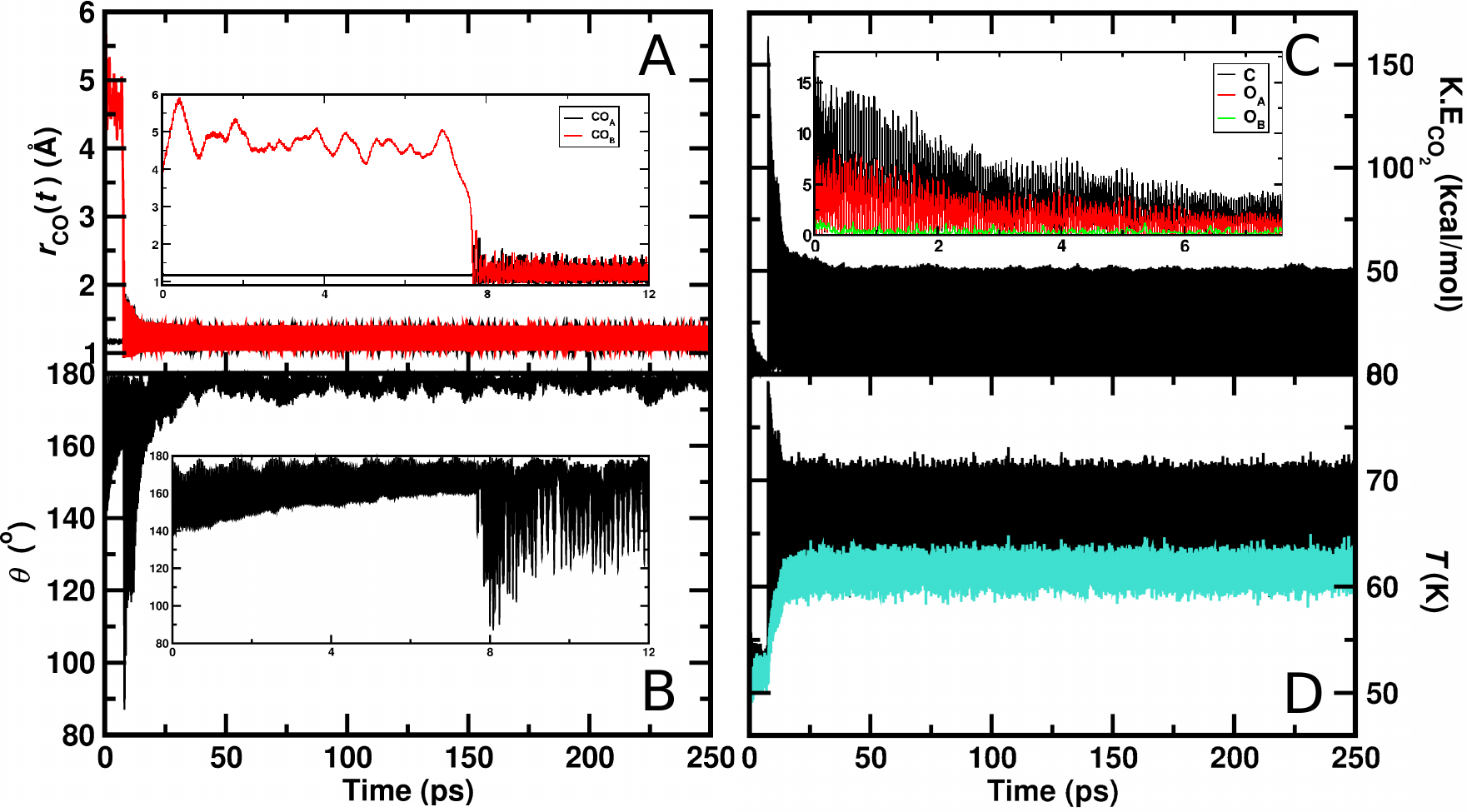}
\caption{Interatomic distances CO$_{\rm A}$ and CO$_{\rm B}$ (panel A)
  and the OCO angle $\theta$ (panel B, running average) from a 250 ps
  simulation using the MMH PES. Initially, $R = 4.66$ \AA\/ and
  $\theta = 135^\circ$. The diffusive motion of O$_{\rm B}$ can be
  seen in the inset before recombination takes place after $\sim 7$
  ps. After recombination both CO stretches are equally highly
  excited. Panel C: Kinetic energy of CO$_2$ along with the
  contribution from each of the atoms. Panel D: Average temperature of
  the ASW (cyan) and the full system (black) before and after CO$_2$
  formation. After recombination there is a steep increase in
  temperature of the entire system (CO$_2$ plus ASW) whereas warming
  of the ASW is more gradual. This suggests that the underlying
  process is vibrational relaxation of hot CO$_2$ on a cool ASW
  surface which gradually warms and assumes a new thermal
  equilibrium.}
\label{fig:tser}
\end{figure}

\noindent
A representative time series for CO${\rm _A}$ + O${\rm _B}$ to form
CO$_2$ is reported in Figure \ref{fig:tser}. From an initial
separation of $R=4.66$ \AA\/ and $\theta = 135^\circ$, CO$_2$ is
formed within $\sim 8$ ps (inset of Figure \ref{fig:tser}A). Upon
recombination, both CO stretch coordinates are highly excited and the
OCO angle fluctuates around $\theta = 180^\circ$ with an amplitude of
$\sim 10^\circ$ (Figures \ref{fig:tser}A and B). For the remaining 250
ps CO$_2$ relaxes slowly and remains as a diffusing product on the ASW
(see also Figure \ref{fig:struc}).\\

\noindent
It is also of interest to follow the kinetic energy and the
temperature of the ASW before and after recombination, see Figures
\ref{fig:tser}C and D. Before CO${\rm _A}$ + O${\rm _B}$ recombination
the average temperature of the water molecules and the entire system
are close to one another and fluctuate around 50 K. Upon recombination
the temperatures as determined from the kinetic energies of the oxygen
and carbon atoms increases considerably by about 10 K on average and
assume different values as the system now consists of a ``hot'' CO$_2$
molecule adsorbed on a cool ASW surface. The temperature of the full
system (recombined CO$_2$ and the ASW) first shows a prominent peak
reaching 75 K right after recombination and relaxing
subsequently. Contrary to that, the temperature of the water molecules
does not show such a spike but rather increases gradually from 50 K to
60 K following relaxation of the CO$_2$ molecule (inset Figure
\ref{fig:tser}D).\\

\begin{figure}
  \centering
  \includegraphics[scale=0.640]{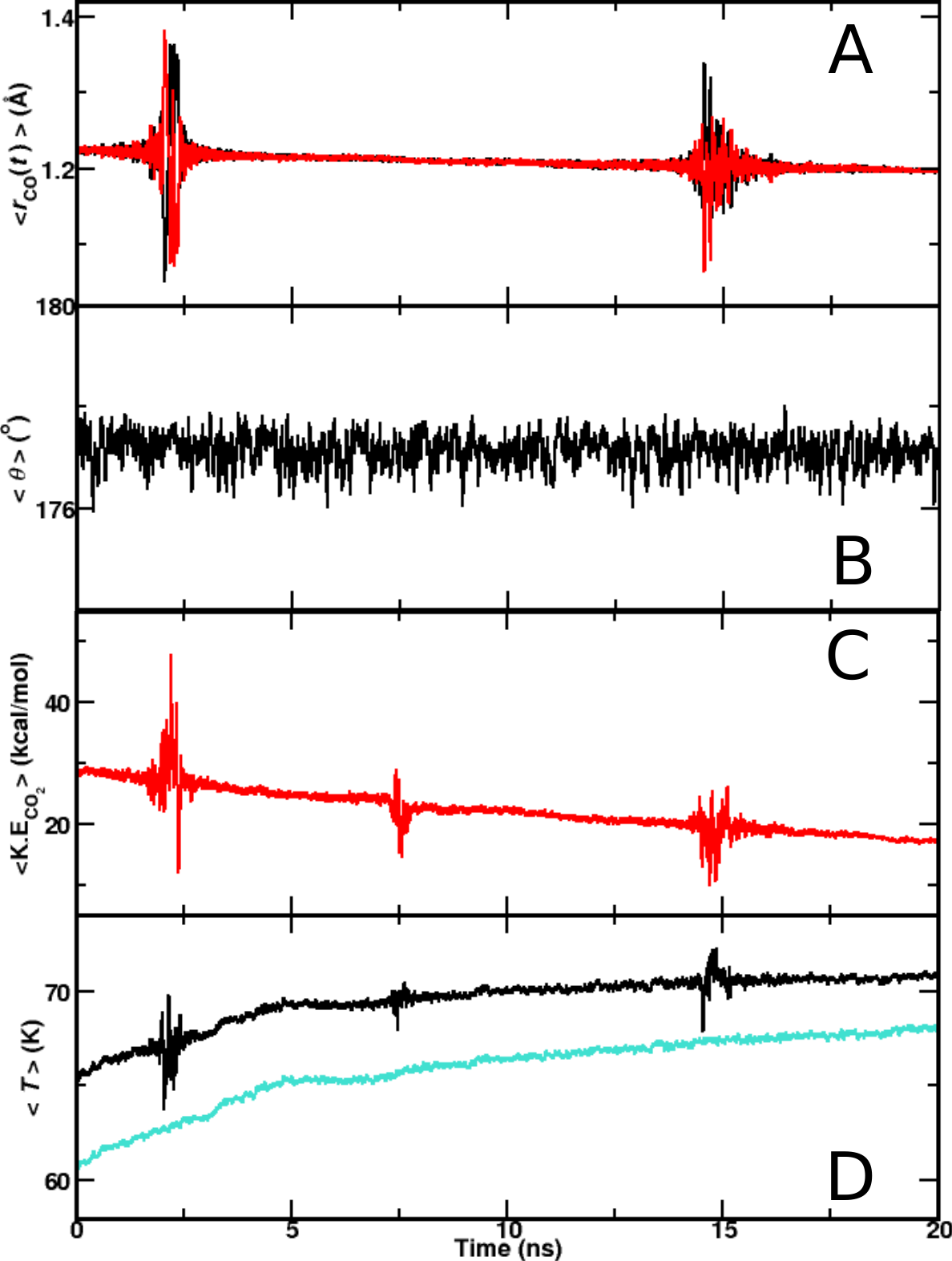}
\caption{Interatomic distances CO${\rm _A}$ and CO${\rm _B}$ (panel
  A), OCO angle (panel B), kinetic energy of CO$_2$ (panel C) and ASW
  (cyan) and overall system (black)(panel D) from 20 ns
  simulation. The spikes are due to collisions of CO$_2$ with the
  surrounding water matrix.}
\label{fig:long.ekin}
\end{figure}

\noindent
To better characterize relaxation of the internal energy, longer (20
ns) simulations were also run. Figure \ref{fig:long.ekin} reports the
CO bond lengths (panel A), the OCO angle (panel B), the kinetic energy
of CO$_2$ (panel C), and the temperatures of the total system (panel
D, black) and the ASW (panel D, cyan). During the entire 20 ns slow
relaxation of the CO stretch amplitudes takes place, interrupted by
occasional large amplitude motions, caused by scattering of the CO$_2$
from collisions with the water molecules. Such collisions are
accompanied by increase of the kinetic energy of the entire CO$_2$
molecule. The raw data for $\theta(t)$ is reported in Figure
S3. Energy exchange between the vibrationally
excited CO$_2$ molecule and the ASW surface continues out to 20 ns and
beyond which is reflected in the continued warming of the water
molecules. However, the slope of the cyan curve in Figure
\ref{fig:long.ekin}D is steeper during the first 5 ns and then
flattens out for the remainder of this trajectory.\\

\noindent
The MMH simulations discussed so far demonstrate that CO+O collisions
lead to formation of CO$_2$ which relaxes on time scales longer than
nanoseconds and does not desorb from the ASW surface.\\

\begin{figure}[H]
  \centering \includegraphics[scale=0.4]{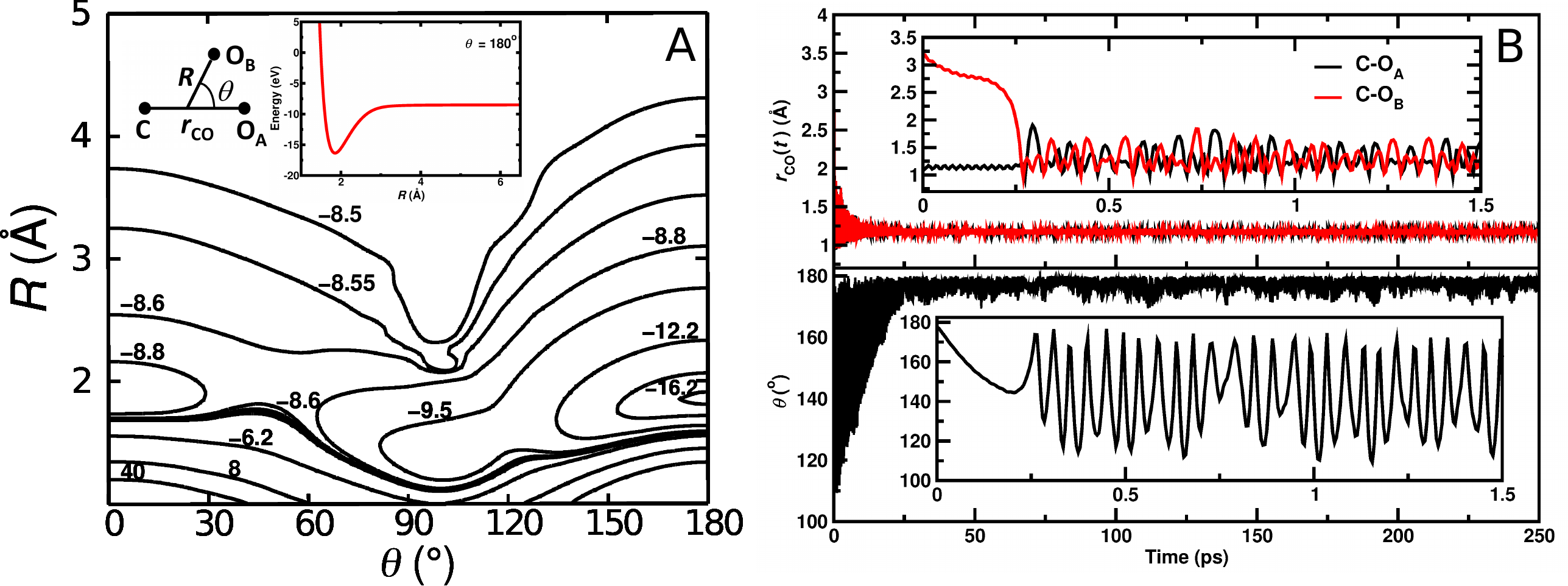}
\caption{Panel A: Two-dimensional cut through the 3d RKHS PES at the
  CCSD(T)-F12 level of theory for the CO + O channel at fixed $r_{\rm
    CO_{\rm A}} = 1.21 $ \AA\/. The inset reports the 1d cut along
  $\theta = 180^\circ$ for $r_{\rm CO_{\rm A}} = 1.21 $ \AA\/. All
  energies in eV and the zero of energy at the dissociation into
  atomic fragments. Panel B: Interatomic distances CO$_{\rm A}$ and
  CO$_{\rm B}$ (top panel, for labels see panel A) and the OCO angle
  (bottom panel) from a 250 ps simulation with $R = 3.9$ \AA\/ and
  $\theta = 180^\circ$ using the RKHS PES. Initially, the CO$_{\rm A}$
  bond (black) is fluctuating around its thermal (50 K) equilibrium
  separation but the amplitude increases considerably after
  recombination due to internal vibrational relaxation. The OCO angle
  relaxes to a quasi-linear structure on the 25 ps time scale.}
\label{fig:rkhs}
\end{figure}

\noindent
Corresponding simulations were carried out with the more realistic but
computationally more expensive RKHS\cite{MM.rkhs:2017} representation
of the CCSD(T)-F12 PES (see SI). Figure \ref{fig:rkhs} reports the two
dimensional PES for $r_{\textrm{CO$_{\rm A}$}}=1.21$ \AA\/ and varying
$R$ and $\theta$, and shows a deep minimum for the linear OCO
structure ($\theta = 180^\circ$) together with the high-energy,
metastable COO structure ($\theta = 0^\circ$), 167.7 kcal/mol higher
in energy than the global minimum. Such a COO intermediate has been
proposed from the interpretation of the C+O$_2$
reaction\cite{dubrin:1964} and was also found in multiconfiguration
SCF\cite{Xantheas1994} and MRCI+Q calculations.\cite{MM.co2:2021} The
two structures are separated by a barrier of 9.55 kcal/mol in going
from the COO to the OCO structure.\\

\noindent
A typical reactive trajectory to form CO$_2$ using the RKHS PES is
reported in Figure \ref{fig:rkhs}B. Again, recombination leads to a
highly vibrationally excited CO$_2$ molecule with rapidly decaying CO
stretch amplitude for the CO-bond that is newly formed (red trace)
whereas the CO$_{\rm A}$ vibrates around 1 \AA\/ initially and extends
to $\sim 1.2$ \AA\/ upon recombination. The OCO angle relaxes to a
quasi-linear structure on the 25 ps time scale, see lower panel in
Figure \ref{fig:rkhs}B. As this 3-dimensional PES also supports a COO
conformation, it is of interest to start simulations from initial
orientations $\theta = 0^\circ$. Indeed, COO formation and
stabilization on the sub-nanosecond time scale is observed, see Figure
S4. Such an intermediate is of potential interest as it
can lead to C + O$_2$ formation.\\

\noindent
The possible final states after collision include (I) CO${\rm _A}$ +
O${\rm _B}$ $\longrightarrow$ CO${\rm _A}$ + O${\rm _B}$ (flyby) (II)
CO${\rm _A}$ + O${\rm _B}$ $\longrightarrow$ CO${\rm _B}$ + O${\rm
  _A}$ (atom exchange) (III) CO${\rm _A}$ + O${\rm _B}$
$\longrightarrow$ CO${\rm _2}$ (CO$_2$ formation) (IV) CO$_{\rm A}$
remains on the surface and O$_{\rm B}$ desorbs (elastic scattering
(ES1)), (V) CO$_{\rm A}$ desorbs and O$_{\rm B}$ remains on the
surface (ES2), and (VI) both, CO and O desorb from the ASW
surface. For each of the initial conditions between 100 and 1000
independent trajectories were run, see Tables S1,
S2 and S3. In total, 32500 simulations
were run to determine the probability and rate for CO$_2$ formation
depending on the initial conditions. All processes I to VI were
observed in the simulations described in the following. One of the
possible processes that was not observed on the nanosecond time scale
is CO$_2$ formation with subsequent desorption of CO$_2$. This is
reminiscent of the situation for O$_2$ formation for which desorption
of molecular oxygen after oxygen atom recombination did not
occur.\cite{MM.oxy:2019}\\

\begin{figure}[H]
\centering \includegraphics[scale=0.185]{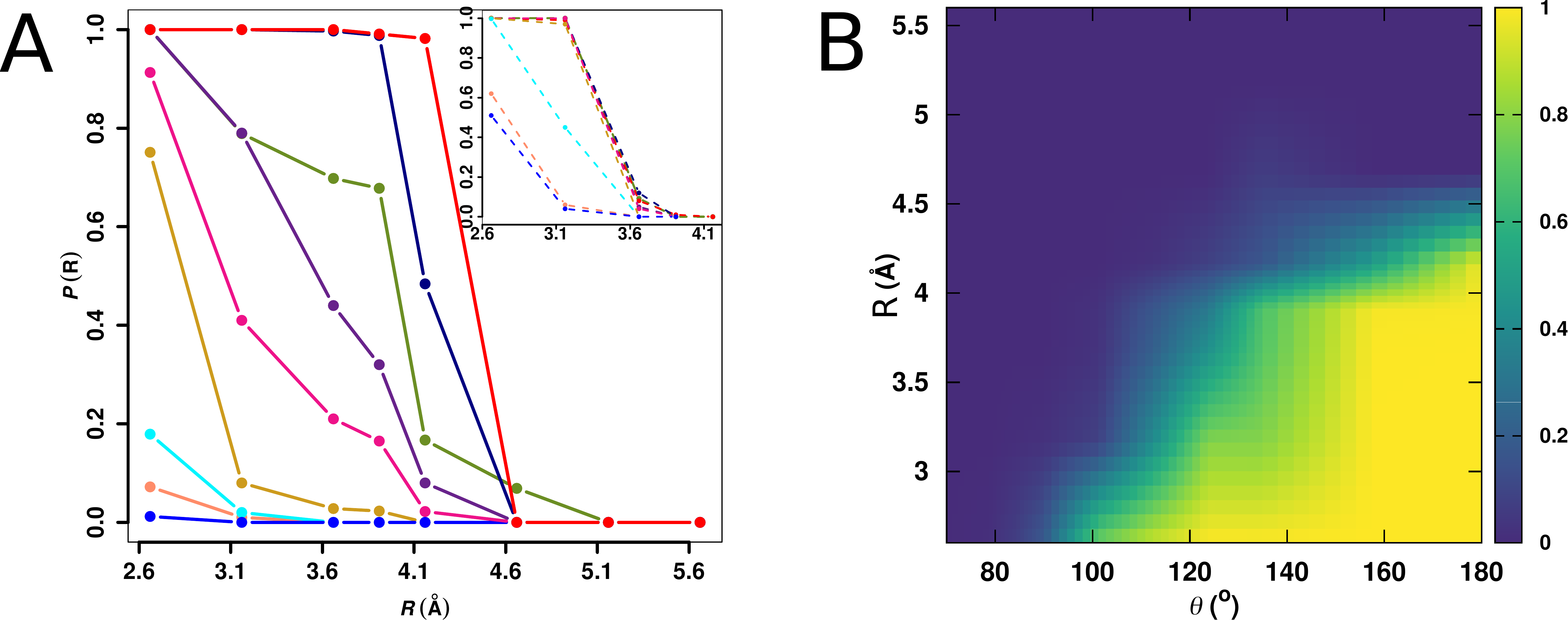}
\caption{Panel A: The CO$_2$ formation probability as a function of
  $R$ and $\theta$ normalized to the number of trajectories with Morse
  potential (main panel) and RKHS (inset) where different colors refer
  to probability for different angles. Color code: darkblue:
  $78.75^\circ$; salmon: $84.375^\circ$; skyblue: $90.0^\circ$, gold:
  $101.25^\circ$, pink: $112.5^\circ$, darkorchid: $123.75^\circ$,
  olive: $135.0^\circ$, navy: $157.5^\circ$, red: $180.0^\circ$. Panel
  B: Plot for the CO$_2$ formation probability as a function of $R$
  and $\theta$ using MMH PES. For a smoother shape or $P(R,\theta)$
  kernel density estimation (KDE)\cite{rosenblatt:1956} was used.}
\label{fig:prob1d}
\end{figure}

\noindent
As the main focus of the present work concerns a) whether or not
CO$_2$ is formed, b) whether CO$_2$ stabilizes and c) what other
processes follow the recombination reaction, the initial separations
between the two reaction partners to explore these questions was
limited to $R \sim 6$ \AA\/. As already found for the
O($^3$P)--O($^3$P) reaction even initial separations of $\sim 10$
\AA\/ can lead to recombination, see Figure S4. However,
to maintain simulation times manageable for this two-dimensional
problem, the initial separations were limited to a somewhat shorter
value of $R$.\\

\noindent
Figure \ref{fig:prob1d}A reports the reaction probability for CO$_2$
formation depending on the initial reaction geometry $(R,\theta)$ for
the MMH (main panel) and the RKHS PES (inset). Simulations starting
from $\theta = 180^\circ$, i.e. along the O$_{\rm A}$C--O$_{\rm B}$
approach have unit reaction probability with initial separations
ranging from $R=2.6$ \AA\/ up to $R = 4.5$ \AA\/. Such separations
cover both situations, the reacting partners within the same and in
two different wells. Moving along $\theta$ it is found that on the MMH
PES the reaction probabilities rapidly decrease whereas on the RKHS
PES they remain close to 1 up to $\theta \approx 100^\circ$ after
which they decrease (inset Figure \ref{fig:prob1d}A). These
differences are due to the different topographies of the two PESs.\\

\noindent
For the MMH model there is a sufficient number of trajectories on a
fine and extensive grid of initial collision geometries to also
construct a 2-dimensional probability function for the reaction
probability, see Figure \ref{fig:prob1d}B. This was done by using
kernel density estimation\cite{rosenblatt:1956} applied to the raw
data from Figure \ref{fig:prob1d}A. It is found that as soon as the
oxygen atom diffuses within a range of $\sim 4$ \AA\/ of the center of
mass of CO molecule, formation of CO$_2$ is highly probable.\\

\begin{figure}[H]
\centering \includegraphics[scale=0.60]{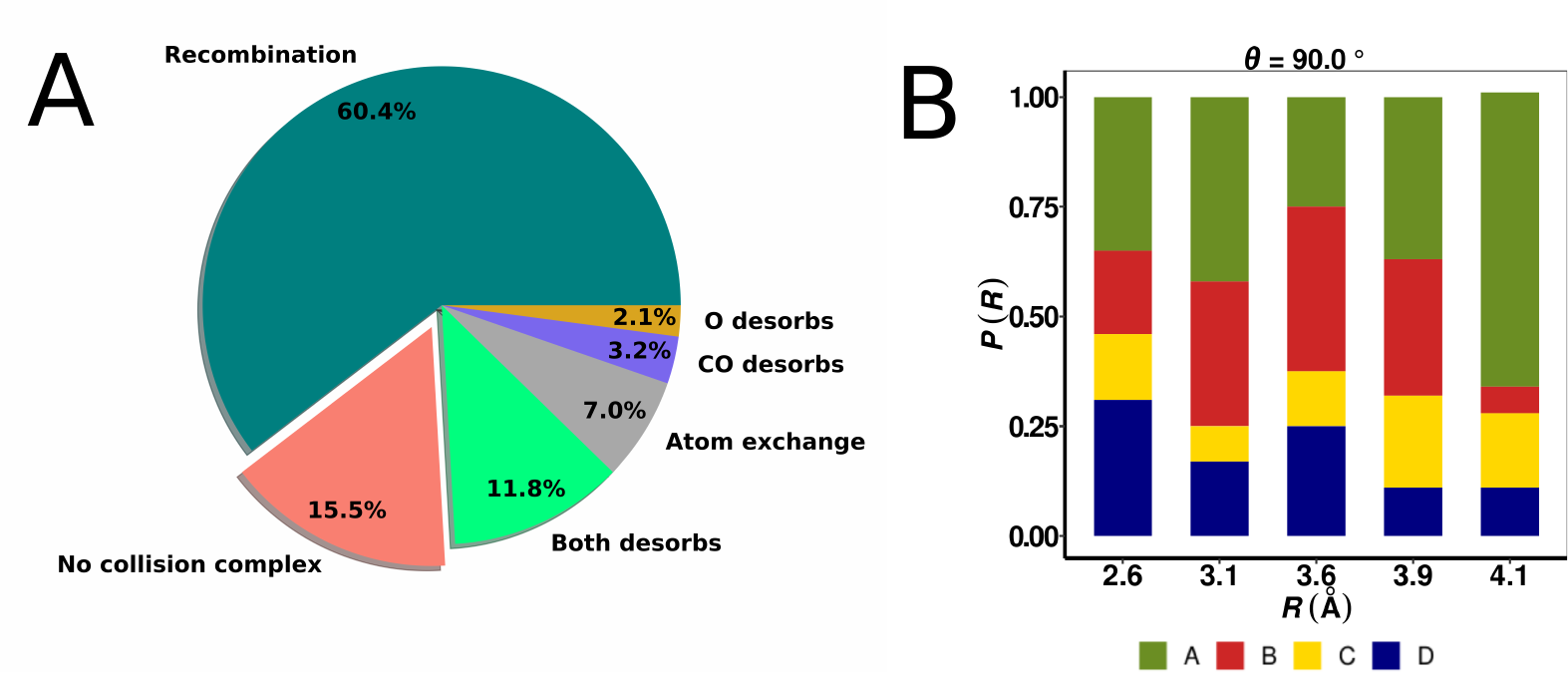}
\caption{Panel A: Classification of outcomes from all the $> 30000$
  simulations using the MMH PES. In more than half the cases CO$_2$ is
  formed whereas for 15 \% of the cases the two collision partners do
  not meet. The remaining 25 \% include atom exchange reactions,
  elastic scattering or desorption of both collision partners. Panel
  B: After atom exchange, there are 4 further possibilities i.e (A)
  both CO and O remain on the water surface, (B) both CO and O desorb,
  (C) O remains and CO desorbs, and (D) CO remains and O desorbs from
  the water surface for $\theta = 90^\circ$. See Figure
    S5 for different angles.}
\label{fig:alloutcomes}
\end{figure}

\noindent
With respect to the distribution of the final state channels, more
than 50 \% of the trajectories run on the MMH PES lead to CO$_2$ and
remain in this state, see Figure \ref{fig:alloutcomes}A. The multiple
long-time (20 ns) simulations indicate that once formed, CO$_2$ is
expected to remain adsorbed on the ASW and vibrationally cools on
considerably longer time scales ($\mu$s to ms). Among the other,
minor, channels, the one not forming a collision complex is most
probable (16 \%), followed by desorption of both reaction partners (12
\%), and the atom exchange reaction (7 \%), whereas desorption of
either CO or O occur rarely.\\

\noindent
The atom exchange reaction is of particular interest because it
provides another recombination channel, but probably on longer
simulation times, if both reaction partners remain on the ASW. Figure
\ref{fig:alloutcomes}B shows that CO$_{\rm B}$ and O$_{\rm A}$
remaining on the ASW after atom exchange (green) is most probable
irrespective of the initial separation between CO$_{\rm A}$ and
O$_{\rm B}$. This is followed by desorption of both, whereas
desorption of either CO$_{\rm B}$ or O$_{\rm A}$ is less likely on
average. Thus, even if the initial collision process leads to atom
exchange, there is still probability for subsequent CO$_2$ formation
because both reaction partners remain on the ASW.\\

\noindent
The present work demonstrates that CO$_2$ formation and stabilization
from collision of CO and O($^1$D) on ASW is possible and occurs on the
sub-nanosecond time scale for the reaction partners within 10
\AA\/. This compares with a radiative lifetime of 110
minutes\cite{garstang:1951} of O($^1$D) which can be formed from
photolysis of H$_2$O\cite{klemm:1975}. Relaxation of the highly
excited CO$_2$ product occurs on much longer time scales which is not
covered here. From previous work on O($^3$P)+O($^3$P) to form O$_2$ it
is known that full vibrational relaxation occurs on the $\mu$s time
scale.\cite{art4,MM.oxy:2019} Similarly, it is expected that CO$_2$
vibrational relaxation extends out to considerably longer time scales
than those covered in the present work (tens of ns). Using a flexible
water model may provide additional channels to speed up vibrational
relaxation of CO$_2$, as was recently reported for O$_2$ on
ASW.\cite{MM.oxy:2019} A fluctuating charge model to capture charge
flow between the initial O($^1$D)+CO($^1 \Sigma^+$) and the CO$_2$
final state is expected to lead to more rapid diffusion of the CO
molecule and hence increased recombination rates (see SI).\\

\noindent
Effects due to quantum nuclear dynamics are not included here but are
expected to be small given the large mass of the particles
involved. In fact, earlier work\cite{MM.oxy:2018} has demonstrated
that the experimentally\cite{oxy.diff.minissale:2013} determined
$T-$dependent mobility of O($^3$P) is a consequence of ASW surface
roughness rather than tunneling at low temperature. Also, the validity
of QCT-based simulations has been explicitly validated for the [CNO]
reactive system by comparing final vibrational state distributions
with time-independent quantum simulations.\cite{MM.cno:2018} The
efficiency of non-adiabatic effects due to O($^1$D) to O($^3$P)
conversion have been estimated\cite{MM.co2:2021} to be 10 \% at 300 K
based on earlier measurements.\cite{Davidson1978} This may change at
lower temperature, though, and explicit inclusion of nonadiabatic
processes for the O($^1$D)+CO association reaction to form CO$_2$
appears to be of interest in the future.\\

\noindent
The simulations on the RKHS PES support the possibility to form the
COO isomer. Early emission spectra found that the C($^{3}$P) +
O$_{2}$($^3\Sigma_{g}^{-}$) reaction generates CO with $v' = 17$ and
that the transition state has the COO configuration rather than
OCO.\cite{thrush:1973} As both, the high-level CCSD(T)-F12 PES and a
recently determined MRCISD+Q/aug-cc-pVTZ ground state
PES\cite{MM.co2:2021} support stabilization of the COO structure, the
present results also indicate that association of CO with O($^1$D) can
lead to formation and stabilization of COO.\\

\noindent
Complementary to the O($^1$D)+CO($^1 \Sigma^+$) reaction considered
here, it is expected that the O($^3$P)+CO($^1 \Sigma^+$) process to
yield triplet (excited) CO$_2$ is also feasible. The O($^3$P)+CO($^1
\Sigma^+$) asymptotic state connects to the excited $^3$A' and $^3$A''
states of CO$_2$ both exhibiting entrance barriers of 0.2 eV and 0.3
eV in the gas phase, respectively, at the CASSCF-MP2 level of
theory\cite{duff:2000} which change to 0.3 eV and 0.4 eV from MRCI+Q
calculations.\cite{MM.co2:2021} Experiments on ASW indicate that the
entrance barrier reduces to between 0.014 and 0.067
eV.\cite{dulieu:2013} Furthermore, symmetry breaking due to the water
environment\cite{pimentel:1979} makes the transition from triplet to
singlet (ground state) CO$_2$ possible. For the O($^3$P)+CO($^1
\Sigma^+$) reaction equally accurate PESs are
available\cite{MM.co2:2021} which allow to follow this pathway at
molecular detail.\\

\noindent
Here, the reaction probability depends on both, the separation and
orientation of the reacting partners. This differs from
O($^3$P)+O($^3$P) recombination to form O$_2$ and increases the
conformational space to be sampled. Thus, initial separations were
limited to $\sim 6$ \AA\/ which, however, does not affect the
generality of the present results. In fact, reactive trajectories were
shown to still sample considerably larger separations ($R \sim 10$
\AA\/) due to surface diffusion of the reactants before CO$_2$
formation occurred. It will be of interest to compare the current
results with those from laboratory experiments along similar lines as
was possible for oxygen atom
recombination.\cite{oxy.diff.minissale:2013} As an example,
experiments involving O($^1$D) have been recently carried out with
acetylene\cite{yan:2020} and formation of CO$_2$ involving O($^1$D) on
mixed CO/H$_2$O films has been reported to be the dominant reaction
channel.\\

\noindent
This work demonstrates that genesis of small polyatomic molecules can
be investigated from computations using high-level electronic
structure and molecular dynamics methods. For the specific case of
CO$_2$ formation along the O($^1$D)+CO($^1 \Sigma^+$) asymptote the
product stabilizes through intermolecular vibrational relaxation (IVR)
and does not desorb from the ASW on the $< 100$ ns time scale. The
fact that O($^1$D) with a radiative lifetime\cite{garstang:1951} of
110 min is considerably more mobile than ground state
O($^3$P)\cite{apkarian:1993,apkarian:1999} should make this process to
form CO$_2$ feasible. Furthermore, formation of the COO isomer is
consistent with early gas-phase experiments. With recent advances in
machine learning of reactive force fields for polyatomic
molecules,\cite{MM.nn:2018,MM.physnet:2019,unke:2021} explicit
characterization of a wide range of molecule formation processes at
low temperatures and under realistic conditions in interstellar space
become possible. Such studies will ideally complement laboratory-based
and observational astrochemistry efforts to better understand molecule
abundances and the chemical evolution of molecular clouds.\\

\section*{Supporting Information}
The supporting information reports Methods, Tables with a summary of
all reactive simulations, and supporting figures for the present
work.\\

\section*{Data Availability Statement}
The data needed for the RKHS representation of the CCSD(T)-F12 PES is
available at \url{https://github.com/MeuwlyGroup/co2.asw}.

\section*{Acknowledgments}
This work was supported by the Swiss National Science Foundation
grants 200021-117810, 200020-188724, the NCCR MUST, and the University
of Basel which is gratefully acknowledged.

\bibliography{astro2}

\providecommand{\latin}[1]{#1}
\makeatletter
\providecommand{\doi}
  {\begingroup\let\do\@makeother\dospecials
  \catcode`\{=1 \catcode`\}=2 \doi@aux}
\providecommand{\doi@aux}[1]{\endgroup\texttt{#1}}
\makeatother
\providecommand*\mcitethebibliography{\thebibliography}
\csname @ifundefined\endcsname{endmcitethebibliography}
  {\let\endmcitethebibliography\endthebibliography}{}
\begin{mcitethebibliography}{37}
\providecommand*\natexlab[1]{#1}
\providecommand*\mciteSetBstSublistMode[1]{}
\providecommand*\mciteSetBstMaxWidthForm[2]{}
\providecommand*\mciteBstWouldAddEndPuncttrue
  {\def\EndOfBibitem{\unskip.}}
\providecommand*\mciteBstWouldAddEndPunctfalse
  {\let\EndOfBibitem\relax}
\providecommand*\mciteSetBstMidEndSepPunct[3]{}
\providecommand*\mciteSetBstSublistLabelBeginEnd[3]{}
\providecommand*\EndOfBibitem{}
\mciteSetBstSublistMode{f}
\mciteSetBstMaxWidthForm{subitem}{(\alph{mcitesubitemcount})}
\mciteSetBstSublistLabelBeginEnd
  {\mcitemaxwidthsubitemform\space}
  {\relax}
  {\relax}

\bibitem[Tielens(2013)]{tielens:2013}
Tielens,~A. G. G.~M. The molecular universe. \emph{Rev. Mod. Phys.}
  \textbf{2013}, \emph{85}, 1021\relax
\mciteBstWouldAddEndPuncttrue
\mciteSetBstMidEndSepPunct{\mcitedefaultmidpunct}
{\mcitedefaultendpunct}{\mcitedefaultseppunct}\relax
\EndOfBibitem
\bibitem[Tielens and Hagen(1982)Tielens, and Hagen]{Tielens82p245}
Tielens,~A. G. G.~M.; Hagen,~W. Model Calculations of the Molecular Composition
  of Interstellar Grain Mantles. \emph{Astron. Astrophys.} \textbf{1982},
  \emph{114}, 245--260\relax
\mciteBstWouldAddEndPuncttrue
\mciteSetBstMidEndSepPunct{\mcitedefaultmidpunct}
{\mcitedefaultendpunct}{\mcitedefaultseppunct}\relax
\EndOfBibitem
\bibitem[Neininger \latin{et~al.}(1998)Neininger, Gu{\'e}lin, Ungerechts,
  Lucas, and Wielebinski]{neininger:1998}
Neininger,~N.; Gu{\'e}lin,~M.; Ungerechts,~H.; Lucas,~R.; Wielebinski,~R.
  Carbon monoxide emission as a precise tracer of molecular gas in the
  Andromeda galaxy. \emph{Nature} \textbf{1998}, \emph{395}, 871--873\relax
\mciteBstWouldAddEndPuncttrue
\mciteSetBstMidEndSepPunct{\mcitedefaultmidpunct}
{\mcitedefaultendpunct}{\mcitedefaultseppunct}\relax
\EndOfBibitem
\bibitem[Whittet \latin{et~al.}(2001)Whittet, Gerakines, Hough, and
  Shenoy]{shenoy:2001}
Whittet,~D.; Gerakines,~P.; Hough,~J.; Shenoy,~S. Interstellar extinction and
  polarization in the Taurus dark clouds: the optical properties of dust near
  the diffuse/dense cloud interface. \emph{Astrophys. J.} \textbf{2001},
  \emph{547}, 872\relax
\mciteBstWouldAddEndPuncttrue
\mciteSetBstMidEndSepPunct{\mcitedefaultmidpunct}
{\mcitedefaultendpunct}{\mcitedefaultseppunct}\relax
\EndOfBibitem
\bibitem[Garrod and Pauly(2011)Garrod, and Pauly]{garrod:2011}
Garrod,~R.~T.; Pauly,~T. On the Formation of CO$_2$ and Other Interstellar
  Ices. \emph{Astrophys. J.} \textbf{2011}, \emph{735}, 15\relax
\mciteBstWouldAddEndPuncttrue
\mciteSetBstMidEndSepPunct{\mcitedefaultmidpunct}
{\mcitedefaultendpunct}{\mcitedefaultseppunct}\relax
\EndOfBibitem
\bibitem[Fournier \latin{et~al.}(1979)Fournier, Deson, Vermeil, and
  Pimentel]{pimentel:1979}
Fournier,~J.; Deson,~J.; Vermeil,~C.; Pimentel,~G. Fluorescence and
  thermoluminescence of N$_2$O, CO, and CO$_2$ in an argon matrix at low
  temperature. \emph{J. Chem. Phys.} \textbf{1979}, \emph{70}, 5726--5730\relax
\mciteBstWouldAddEndPuncttrue
\mciteSetBstMidEndSepPunct{\mcitedefaultmidpunct}
{\mcitedefaultendpunct}{\mcitedefaultseppunct}\relax
\EndOfBibitem
\bibitem[Ruffle and Herbst(2001)Ruffle, and Herbst]{herbst:2001}
Ruffle,~D.~P.; Herbst,~E. New models of interstellar gas-grain chemistry - III.
  Solid CO$_2$. \emph{Mon. Not. R. Astron. Soc.} \textbf{2001}, \emph{324},
  1054--1062\relax
\mciteBstWouldAddEndPuncttrue
\mciteSetBstMidEndSepPunct{\mcitedefaultmidpunct}
{\mcitedefaultendpunct}{\mcitedefaultseppunct}\relax
\EndOfBibitem
\bibitem[Minissale \latin{et~al.}(2013)Minissale, Congiu, Manic{\`o},
  Pirronello, and Dulieu]{dulieu:2013}
Minissale,~M.; Congiu,~E.; Manic{\`o},~G.; Pirronello,~V.; Dulieu,~F. CO$_2$
  formation on interstellar dust grains: a detailed study of the barrier of the
  CO+ O channel. \emph{Astron. Astrophys.} \textbf{2013}, \emph{559}, A49\relax
\mciteBstWouldAddEndPuncttrue
\mciteSetBstMidEndSepPunct{\mcitedefaultmidpunct}
{\mcitedefaultendpunct}{\mcitedefaultseppunct}\relax
\EndOfBibitem
\bibitem[San Vicente~Veliz \latin{et~al.}(2021)San Vicente~Veliz, Koner,
  Schwilk, Bemish, and Meuwly]{MM.co2:2021}
San Vicente~Veliz,~J.~C.; Koner,~D.; Schwilk,~M.; Bemish,~R.~J.; Meuwly,~M. The
  C($^{3}$P) + O$_{2}$($^3\Sigma_{g}^{-}$) $\leftrightarrow$ CO$_2$
  $\leftrightarrow$ CO($^{1}\Sigma^{+}$)+ O($^{1}$D)/O($^{3}$P) Reaction:
  Thermal and Vibrational Relaxation Rates from 15 K to 20000 K. \emph{Phys.
  Chem. Chem. Phys.} \textbf{2021}, \emph{in print}, in print\relax
\mciteBstWouldAddEndPuncttrue
\mciteSetBstMidEndSepPunct{\mcitedefaultmidpunct}
{\mcitedefaultendpunct}{\mcitedefaultseppunct}\relax
\EndOfBibitem
\bibitem[Ioppolo \latin{et~al.}(2011)Ioppolo, Van~Boheemen, Cuppen,
  Van~Dishoeck, and Linnartz]{linnartz:2011}
Ioppolo,~S.; Van~Boheemen,~Y.; Cuppen,~H.; Van~Dishoeck,~E.; Linnartz,~H.
  Surface formation of CO$_2$ ice at low temperatures. \emph{Mon. Not. R.
  Astron. Soc.} \textbf{2011}, \emph{413}, 2281--2287\relax
\mciteBstWouldAddEndPuncttrue
\mciteSetBstMidEndSepPunct{\mcitedefaultmidpunct}
{\mcitedefaultendpunct}{\mcitedefaultseppunct}\relax
\EndOfBibitem
\bibitem[Qasim \latin{et~al.}(2019)Qasim, Lamberts, He, Chuang, Fedoseev,
  Ioppolo, Boogert, and Linnartz]{linnartz:2019}
Qasim,~D.; Lamberts,~T.; He,~J.; Chuang,~K.-J.; Fedoseev,~G.; Ioppolo,~S.;
  Boogert,~A.; Linnartz,~H. Extension of the HCOOH and CO$_2$ solid-state
  reaction network during the CO freeze-out stage: inclusion of H$_2$CO.
  \emph{Astron. Astrophys.} \textbf{2019}, \emph{626}, A118\relax
\mciteBstWouldAddEndPuncttrue
\mciteSetBstMidEndSepPunct{\mcitedefaultmidpunct}
{\mcitedefaultendpunct}{\mcitedefaultseppunct}\relax
\EndOfBibitem
\bibitem[Minissale \latin{et~al.}(2013)Minissale, Congiu, Baouche, Chaabouni,
  Moudens, Dulieu, Accolla, Cazaux, Manico, and
  Pirronello]{oxy.diff.minissale:2013}
Minissale,~M.; Congiu,~E.; Baouche,~S.; Chaabouni,~H.; Moudens,~A.; Dulieu,~F.;
  Accolla,~M.; Cazaux,~S.; Manico,~G.; Pirronello,~V. Quantum Tunneling of
  Oxygen Atoms on Very Cold Surfaces. \emph{Phys. Rev. Lett.} \textbf{2013},
  \emph{111}, 053201\relax
\mciteBstWouldAddEndPuncttrue
\mciteSetBstMidEndSepPunct{\mcitedefaultmidpunct}
{\mcitedefaultendpunct}{\mcitedefaultseppunct}\relax
\EndOfBibitem
\bibitem[Lee and Meuwly(2014)Lee, and Meuwly]{MM.oxy:2014}
Lee,~M.~W.; Meuwly,~M. Diffusion of atomic oxygen relevant to water formation
  in amorphous interstellar ices. \emph{Faraday Discuss.} \textbf{2014},
  \emph{168}, 205--222\relax
\mciteBstWouldAddEndPuncttrue
\mciteSetBstMidEndSepPunct{\mcitedefaultmidpunct}
{\mcitedefaultendpunct}{\mcitedefaultseppunct}\relax
\EndOfBibitem
\bibitem[Pezzella \latin{et~al.}(2018)Pezzella, Unke, and Meuwly]{MM.oxy:2018}
Pezzella,~M.; Unke,~O.~T.; Meuwly,~M. Molecular Oxygen Formation in
  Interstellar Ices Does Not Require Tunneling. \emph{J. Phys. Chem. Lett.}
  \textbf{2018}, \emph{9}, 1822--1826\relax
\mciteBstWouldAddEndPuncttrue
\mciteSetBstMidEndSepPunct{\mcitedefaultmidpunct}
{\mcitedefaultendpunct}{\mcitedefaultseppunct}\relax
\EndOfBibitem
\bibitem[Bergin and Tafalla(2007)Bergin, and Tafalla]{bergin:2007}
Bergin,~E.~A.; Tafalla,~M. Cold dark clouds: the initial conditions for star
  formation. \emph{Annu. Rev. Astron. Astrophys.} \textbf{2007}, \emph{45},
  339--396\relax
\mciteBstWouldAddEndPuncttrue
\mciteSetBstMidEndSepPunct{\mcitedefaultmidpunct}
{\mcitedefaultendpunct}{\mcitedefaultseppunct}\relax
\EndOfBibitem
\bibitem[Danilychev and Apkarian(1993)Danilychev, and Apkarian]{apkarian:1993}
Danilychev,~A.; Apkarian,~V. Temperature induced mobility and recombination of
  atomic oxygen in crystalline Kr and Xe. I. Experiment. \emph{J. Chem. Phys.}
  \textbf{1993}, \emph{99}, 8617--8627\relax
\mciteBstWouldAddEndPuncttrue
\mciteSetBstMidEndSepPunct{\mcitedefaultmidpunct}
{\mcitedefaultendpunct}{\mcitedefaultseppunct}\relax
\EndOfBibitem
\bibitem[Apkarian and Schwentner(1999)Apkarian, and Schwentner]{apkarian:1999}
Apkarian,~V.; Schwentner,~N. Molecular photodynamics in rare gas solids.
  \emph{Chem. Rev.} \textbf{1999}, \emph{99}, 1481--1514\relax
\mciteBstWouldAddEndPuncttrue
\mciteSetBstMidEndSepPunct{\mcitedefaultmidpunct}
{\mcitedefaultendpunct}{\mcitedefaultseppunct}\relax
\EndOfBibitem
\bibitem[Pezzella and Meuwly(2019)Pezzella, and Meuwly]{MM.oxy:2019}
Pezzella,~M.; Meuwly,~M. O$_2$ formation in cold environments. \emph{Phys.
  Chem. Chem. Phys.} \textbf{2019}, \emph{21}, 6247--6255\relax
\mciteBstWouldAddEndPuncttrue
\mciteSetBstMidEndSepPunct{\mcitedefaultmidpunct}
{\mcitedefaultendpunct}{\mcitedefaultseppunct}\relax
\EndOfBibitem
\bibitem[Pezzella \latin{et~al.}(2020)Pezzella, Koner, and Meuwly]{MM.o2:2020}
Pezzella,~M.; Koner,~D.; Meuwly,~M. Formation and Stabilization of Ground and
  Excited-State Singlet O$_2$ upon Recombination of $^3$P Oxygen on Amorphous
  Solid Water. \emph{J. Phys. Chem. Lett.} \textbf{2020}, \emph{11},
  2171--2176\relax
\mciteBstWouldAddEndPuncttrue
\mciteSetBstMidEndSepPunct{\mcitedefaultmidpunct}
{\mcitedefaultendpunct}{\mcitedefaultseppunct}\relax
\EndOfBibitem
\bibitem[Stief \latin{et~al.}(1975)Stief, Payne, and Klemm]{klemm:1975}
Stief,~L.~J.; Payne,~W.~A.; Klemm,~R.~B. A flash photolysis--resonance
  fluorescence study of the formation of O($^1$D) in the photolysis of water
  and the reaction of O($^1$D) with H$_2$, Ar, and He-. \emph{J. Chem. Phys.}
  \textbf{1975}, \emph{62}, 4000--4008\relax
\mciteBstWouldAddEndPuncttrue
\mciteSetBstMidEndSepPunct{\mcitedefaultmidpunct}
{\mcitedefaultendpunct}{\mcitedefaultseppunct}\relax
\EndOfBibitem
\bibitem[Garstang(1951)]{garstang:1951}
Garstang,~R. Energy levels and transition probabilities in p 2 and p 4
  configurations. \emph{Mon. Not. R. Astron. Soc.} \textbf{1951}, \emph{111},
  115--124\relax
\mciteBstWouldAddEndPuncttrue
\mciteSetBstMidEndSepPunct{\mcitedefaultmidpunct}
{\mcitedefaultendpunct}{\mcitedefaultseppunct}\relax
\EndOfBibitem
\bibitem[Schmidt \latin{et~al.}(2019)Schmidt, Swiderek, and
  Bredehöft]{schmidt:2019}
Schmidt,~F.; Swiderek,~P.; Bredehöft,~J.~H. Formation of Formic Acid,
  Formaldehyde, and Carbon Dioxide by Electron-Induced Chemistry in Ices of
  Water and Carbon Monoxide. \emph{ACS Earth Space Chem.} \textbf{2019},
  \emph{3}, 1974--1986\relax
\mciteBstWouldAddEndPuncttrue
\mciteSetBstMidEndSepPunct{\mcitedefaultmidpunct}
{\mcitedefaultendpunct}{\mcitedefaultseppunct}\relax
\EndOfBibitem
\bibitem[Shimanouchi \latin{et~al.}(1978)Shimanouchi, Matsuura, Ogawa, and
  Harada]{shimanouchi:1978}
Shimanouchi,~T.; Matsuura,~H.; Ogawa,~Y.; Harada,~I. Tables of molecular
  vibrational frequencies. \emph{J. Phys. Chem. Ref. Data} \textbf{1978},
  \emph{7}, 1323--1444\relax
\mciteBstWouldAddEndPuncttrue
\mciteSetBstMidEndSepPunct{\mcitedefaultmidpunct}
{\mcitedefaultendpunct}{\mcitedefaultseppunct}\relax
\EndOfBibitem
\bibitem[Unke and Meuwly(2017)Unke, and Meuwly]{MM.rkhs:2017}
Unke,~O.~T.; Meuwly,~M. Toolkit for the Construction of Reproducing
  Kernel-Based Representations of Data: Application to Multidimensional
  Potential Energy Surfaces. \emph{J. Chem. Inf. Model.} \textbf{2017},
  \emph{57}, 1923--1931\relax
\mciteBstWouldAddEndPuncttrue
\mciteSetBstMidEndSepPunct{\mcitedefaultmidpunct}
{\mcitedefaultendpunct}{\mcitedefaultseppunct}\relax
\EndOfBibitem
\bibitem[Dubrin \latin{et~al.}(1964)Dubrin, MacKay, Pandow, and
  Wolfgang]{dubrin:1964}
Dubrin,~J.; MacKay,~C.; Pandow,~M.; Wolfgang,~R. Reactions of atomic carbon
  with $\pi$-bonded inorganic molecules. \emph{J. Inorg. Nuc. Chem.}
  \textbf{1964}, \emph{26}, 2113--2122\relax
\mciteBstWouldAddEndPuncttrue
\mciteSetBstMidEndSepPunct{\mcitedefaultmidpunct}
{\mcitedefaultendpunct}{\mcitedefaultseppunct}\relax
\EndOfBibitem
\bibitem[Xantheas and Ruedenberg(1994)Xantheas, and Ruedenberg]{Xantheas1994}
Xantheas,~S.~S.; Ruedenberg,~K. Potential energy surfaces of carbon dioxide.
  \emph{Int. J. Quant. Chem.} \textbf{1994}, \emph{49}, 409--427\relax
\mciteBstWouldAddEndPuncttrue
\mciteSetBstMidEndSepPunct{\mcitedefaultmidpunct}
{\mcitedefaultendpunct}{\mcitedefaultseppunct}\relax
\EndOfBibitem
\bibitem[Rosenblatt(1956)]{rosenblatt:1956}
Rosenblatt,~M. Remarks on Some Nonparametric Estimates of a Density Function.
  \emph{Ann. Math. Statist.} \textbf{1956}, \emph{27}, 832--837\relax
\mciteBstWouldAddEndPuncttrue
\mciteSetBstMidEndSepPunct{\mcitedefaultmidpunct}
{\mcitedefaultendpunct}{\mcitedefaultseppunct}\relax
\EndOfBibitem
\bibitem[Hama \latin{et~al.}(2010)Hama, Yokoyama, Yabushita, and
  Kawasaki]{art4}
Hama,~T.; Yokoyama,~M.; Yabushita,~A.; Kawasaki,~M. Role of OH radicals in the
  formation of oxygen molecules following vacuum ultraviolet photodissociation
  ofa amorphous solid water. \emph{J. Chem. Phys.} \textbf{2010}, \emph{133},
  104504\relax
\mciteBstWouldAddEndPuncttrue
\mciteSetBstMidEndSepPunct{\mcitedefaultmidpunct}
{\mcitedefaultendpunct}{\mcitedefaultseppunct}\relax
\EndOfBibitem
\bibitem[Koner \latin{et~al.}(2018)Koner, Bemish, and Meuwly]{MM.cno:2018}
Koner,~D.; Bemish,~R.~J.; Meuwly,~M. The C($^3$P) + NO(X$^2\Pi$) $\rightarrow$
  O($^3$P) + CN(X$^2\Sigma^+$), N($^2$D)/N($^4$S) + CO(X$^1\Sigma^+$) reaction:
  Rates, branching ratios, and final states from 15 K to 20 000 K. \emph{J.
  Chem. Phys.} \textbf{2018}, \emph{149}, 094305\relax
\mciteBstWouldAddEndPuncttrue
\mciteSetBstMidEndSepPunct{\mcitedefaultmidpunct}
{\mcitedefaultendpunct}{\mcitedefaultseppunct}\relax
\EndOfBibitem
\bibitem[Davidson \latin{et~al.}(1978)Davidson, Schiff, Brown, and
  Howard]{Davidson1978}
Davidson,~J.~A.; Schiff,~H.~I.; Brown,~T.~J.; Howard,~C.~J. Temperature
  dependence of the deactivation of O($^{1}$D) by CO from 113 to 333 K.
  \emph{J. Chem. Phys.} \textbf{1978}, \emph{69}, 1216--1217\relax
\mciteBstWouldAddEndPuncttrue
\mciteSetBstMidEndSepPunct{\mcitedefaultmidpunct}
{\mcitedefaultendpunct}{\mcitedefaultseppunct}\relax
\EndOfBibitem
\bibitem[Ogryzlo \latin{et~al.}(1973)Ogryzlo, Reilly, and Thrush]{thrush:1973}
Ogryzlo,~E.; Reilly,~J.; Thrush,~B. Vibrational excitation of CO from the
  reaction C+ O$_2$. \emph{Chem. Phys. Lett.} \textbf{1973}, \emph{23},
  37--39\relax
\mciteBstWouldAddEndPuncttrue
\mciteSetBstMidEndSepPunct{\mcitedefaultmidpunct}
{\mcitedefaultendpunct}{\mcitedefaultseppunct}\relax
\EndOfBibitem
\bibitem[Braunstein and Duff(2000)Braunstein, and Duff]{duff:2000}
Braunstein,~M.; Duff,~J.~W. Electronic structure and dynamics of
  O($^3$P)+CO($^1 \Sigma^+$) collisions. \emph{J. Chem. Phys.} \textbf{2000},
  \emph{112}, 2736--2745\relax
\mciteBstWouldAddEndPuncttrue
\mciteSetBstMidEndSepPunct{\mcitedefaultmidpunct}
{\mcitedefaultendpunct}{\mcitedefaultseppunct}\relax
\EndOfBibitem
\bibitem[Yan \latin{et~al.}(2020)Yan, Teng, Chen, Zhong, Rousso, Zhao, Ma,
  Wysocki, and Ju]{yan:2020}
Yan,~C.; Teng,~C.~C.; Chen,~T.; Zhong,~H.; Rousso,~A.; Zhao,~H.; Ma,~G.;
  Wysocki,~G.; Ju,~Y. The kinetic study of excited singlet oxygen atom O (1D)
  reactions with acetylene. \emph{Comb. Flame} \textbf{2020}, \emph{212},
  135--141\relax
\mciteBstWouldAddEndPuncttrue
\mciteSetBstMidEndSepPunct{\mcitedefaultmidpunct}
{\mcitedefaultendpunct}{\mcitedefaultseppunct}\relax
\EndOfBibitem
\bibitem[Unke and Meuwly(2018)Unke, and Meuwly]{MM.nn:2018}
Unke,~O.~T.; Meuwly,~M. A reactive, scalable, and transferable model for
  molecular energies from a neural network approach based on local information.
  \emph{J. Chem. Phys.} \textbf{2018}, \emph{148}, 241708\relax
\mciteBstWouldAddEndPuncttrue
\mciteSetBstMidEndSepPunct{\mcitedefaultmidpunct}
{\mcitedefaultendpunct}{\mcitedefaultseppunct}\relax
\EndOfBibitem
\bibitem[Unke and Meuwly(2019)Unke, and Meuwly]{MM.physnet:2019}
Unke,~O.~T.; Meuwly,~M. PhysNet: A neural network for predicting energies,
  forces, dipole moments, and partial charges. \emph{J. Chem. Theo. Comp.}
  \textbf{2019}, \emph{15}, 3678--3693\relax
\mciteBstWouldAddEndPuncttrue
\mciteSetBstMidEndSepPunct{\mcitedefaultmidpunct}
{\mcitedefaultendpunct}{\mcitedefaultseppunct}\relax
\EndOfBibitem
\bibitem[Unke \latin{et~al.}(2021)Unke, Chmiela, Sauceda, Gastegger, Poltavsky,
  Sch{\"u}tt, Tkatchenko, and M{\"u}ller]{unke:2021}
Unke,~O.~T.; Chmiela,~S.; Sauceda,~H.~E.; Gastegger,~M.; Poltavsky,~I.;
  Sch{\"u}tt,~K.~T.; Tkatchenko,~A.; M{\"u}ller,~K.-R. Machine learning force
  fields. \emph{J. Chem. Theo. Comp.} \textbf{2021}, \relax
\mciteBstWouldAddEndPunctfalse
\mciteSetBstMidEndSepPunct{\mcitedefaultmidpunct}
{}{\mcitedefaultseppunct}\relax
\EndOfBibitem
\end{mcitethebibliography}


\providecommand{\latin}[1]{#1}
\makeatletter
\providecommand{\doi}
  {\begingroup\let\do\@makeother\dospecials
  \catcode`\{=1 \catcode`\}=2 \doi@aux}
\providecommand{\doi@aux}[1]{\endgroup\texttt{#1}}
\makeatother
\providecommand*\mcitethebibliography{\thebibliography}
\csname @ifundefined\endcsname{endmcitethebibliography}
  {\let\endmcitethebibliography\endthebibliography}{}
\begin{mcitethebibliography}{17}
\providecommand*\natexlab[1]{#1}
\providecommand*\mciteSetBstSublistMode[1]{}
\providecommand*\mciteSetBstMaxWidthForm[2]{}
\providecommand*\mciteBstWouldAddEndPuncttrue
  {\def\EndOfBibitem{\unskip.}}
\providecommand*\mciteBstWouldAddEndPunctfalse
  {\let\EndOfBibitem\relax}
\providecommand*\mciteSetBstMidEndSepPunct[3]{}
\providecommand*\mciteSetBstSublistLabelBeginEnd[3]{}
\providecommand*\EndOfBibitem{}
\mciteSetBstSublistMode{f}
\mciteSetBstMaxWidthForm{subitem}{(\alph{mcitesubitemcount})}
\mciteSetBstSublistLabelBeginEnd
  {\mcitemaxwidthsubitemform\space}
  {\relax}
  {\relax}

\bibitem[T.~Nagy and Meuwly(2014)T.~Nagy, and Meuwly]{msarmd}
T.~Nagy,~J. Y.~R.; Meuwly,~M. Multi-Surface Adiabatic Reactive Molecular
  Dynamics. \emph{J. Chem. Theo. Comp.} \textbf{2014}, \emph{10},
  1366--1375\relax
\mciteBstWouldAddEndPuncttrue
\mciteSetBstMidEndSepPunct{\mcitedefaultmidpunct}
{\mcitedefaultendpunct}{\mcitedefaultseppunct}\relax
\EndOfBibitem
\bibitem[Werner \latin{et~al.}(2020)Werner, Knowles, Manby, Black, Doll,
  Hesselmann, Kats, Koehn, Korona, Kreplin, Ma, Miller, Mitrushchenkov,
  Peterson, Polyak, Rauhut, and Sibaev]{molpro:2020}
Werner,~H.-J. \latin{et~al.}  The Molpro quantum chemistry package. \emph{J.
  Chem. Phys.} \textbf{2020}, \emph{152}\relax
\mciteBstWouldAddEndPuncttrue
\mciteSetBstMidEndSepPunct{\mcitedefaultmidpunct}
{\mcitedefaultendpunct}{\mcitedefaultseppunct}\relax
\EndOfBibitem
\bibitem[Werner and Knowles(1988)Werner, and Knowles]{werner1988efficient}
Werner,~H.-J.; Knowles,~P.~J. An efficient internally contracted
  multiconfiguration--reference configuration interaction method. \emph{J.
  Chem. Phys.} \textbf{1988}, \emph{89}, 5803--5814\relax
\mciteBstWouldAddEndPuncttrue
\mciteSetBstMidEndSepPunct{\mcitedefaultmidpunct}
{\mcitedefaultendpunct}{\mcitedefaultseppunct}\relax
\EndOfBibitem
\bibitem[Dunning~Jr(1989)]{dunning1989gaussian}
Dunning~Jr,~T.~H. Gaussian basis sets for use in correlated molecular
  calculations. I. The atoms boron through neon and hydrogen. \emph{J. Chem.
  Phys.} \textbf{1989}, \emph{90}, 1007--1023\relax
\mciteBstWouldAddEndPuncttrue
\mciteSetBstMidEndSepPunct{\mcitedefaultmidpunct}
{\mcitedefaultendpunct}{\mcitedefaultseppunct}\relax
\EndOfBibitem
\bibitem[Best \latin{et~al.}(2012)Best, Zhu, Shim, Lopes, Mittal, Feig, and
  MacKerell~Jr]{best2012optimization}
Best,~R.~B.; Zhu,~X.; Shim,~J.; Lopes,~P.~E.; Mittal,~J.; Feig,~M.;
  MacKerell~Jr,~A.~D. Optimization of the additive CHARMM all-atom protein
  force field targeting improved sampling of the backbone $\phi$, $\psi$ and
  side-chain $\chi$1 and $\chi$2 dihedral angles. \emph{J. Chem. Theo. Comp.}
  \textbf{2012}, \emph{8}, 3257--3273\relax
\mciteBstWouldAddEndPuncttrue
\mciteSetBstMidEndSepPunct{\mcitedefaultmidpunct}
{\mcitedefaultendpunct}{\mcitedefaultseppunct}\relax
\EndOfBibitem
\bibitem[Pezzella and Meuwly(2019)Pezzella, and Meuwly]{MM.oxy:2019}
Pezzella,~M.; Meuwly,~M. O$_2$ formation in cold environments. \emph{Phys.
  Chem. Chem. Phys.} \textbf{2019}, \emph{21}, 6247--6255\relax
\mciteBstWouldAddEndPuncttrue
\mciteSetBstMidEndSepPunct{\mcitedefaultmidpunct}
{\mcitedefaultendpunct}{\mcitedefaultseppunct}\relax
\EndOfBibitem
\bibitem[Pezzella \latin{et~al.}(2020)Pezzella, Koner, and Meuwly]{MM.o2:2020}
Pezzella,~M.; Koner,~D.; Meuwly,~M. Formation and Stabilization of Ground and
  Excited-State Singlet O$_2$ upon Recombination of $^3$P Oxygen on Amorphous
  Solid Water. \emph{J. Phys. Chem. Lett.} \textbf{2020}, \emph{11},
  2171--2176\relax
\mciteBstWouldAddEndPuncttrue
\mciteSetBstMidEndSepPunct{\mcitedefaultmidpunct}
{\mcitedefaultendpunct}{\mcitedefaultseppunct}\relax
\EndOfBibitem
\bibitem[Adler \latin{et~al.}(2007)Adler, Knizia, and Werner]{adler2007simple}
Adler,~T.~B.; Knizia,~G.; Werner,~H.-J. A simple and efficient CCSD(T)-F12
  approximation. 2007\relax
\mciteBstWouldAddEndPuncttrue
\mciteSetBstMidEndSepPunct{\mcitedefaultmidpunct}
{\mcitedefaultendpunct}{\mcitedefaultseppunct}\relax
\EndOfBibitem
\bibitem[Peterson \latin{et~al.}(2008)Peterson, Adler, and
  Werner]{peterson2008systematically}
Peterson,~K.~A.; Adler,~T.~B.; Werner,~H.-J. Systematically convergent basis
  sets for explicitly correlated wavefunctions: The atoms H, He, B--Ne, and
  Al--Ar. \emph{J. Chem. Phys.} \textbf{2008}, \emph{128}, 084102\relax
\mciteBstWouldAddEndPuncttrue
\mciteSetBstMidEndSepPunct{\mcitedefaultmidpunct}
{\mcitedefaultendpunct}{\mcitedefaultseppunct}\relax
\EndOfBibitem
\bibitem[Unke and Meuwly(2017)Unke, and Meuwly]{MM.rkhs:2017}
Unke,~O.~T.; Meuwly,~M. Toolkit for the Construction of Reproducing
  Kernel-Based Representations of Data: Application to Multidimensional
  Potential Energy Surfaces. \emph{J. Chem. Inf. Model.} \textbf{2017},
  \emph{57}, 1923--1931\relax
\mciteBstWouldAddEndPuncttrue
\mciteSetBstMidEndSepPunct{\mcitedefaultmidpunct}
{\mcitedefaultendpunct}{\mcitedefaultseppunct}\relax
\EndOfBibitem
\bibitem[Jorgensen \latin{et~al.}(1983)Jorgensen, Chandrasekhar, Madura, Impey,
  and Klein]{jorgensen-tip3p}
Jorgensen,~W.~L.; Chandrasekhar,~J.; Madura,~J.~D.; Impey,~R.~W.; Klein,~M.~L.
  Comparison of simple potential functions for simulating liquid water.
  \emph{J. Chem. Phys.} \textbf{1983}, \emph{79}, 926--935\relax
\mciteBstWouldAddEndPuncttrue
\mciteSetBstMidEndSepPunct{\mcitedefaultmidpunct}
{\mcitedefaultendpunct}{\mcitedefaultseppunct}\relax
\EndOfBibitem
\bibitem[Wang \latin{et~al.}(2013)Wang, Head-Gordon, Ponder, Ren, Chodera,
  Eastman, Martinez, and Pande]{wang2013systematic}
Wang,~L.-P.; Head-Gordon,~T.; Ponder,~J.~W.; Ren,~P.; Chodera,~J.~D.;
  Eastman,~P.~K.; Martinez,~T.~J.; Pande,~V.~S. Systematic improvement of a
  classical molecular model of water. \emph{J. Phys. Chem. B} \textbf{2013},
  \emph{117}, 9956--9972\relax
\mciteBstWouldAddEndPuncttrue
\mciteSetBstMidEndSepPunct{\mcitedefaultmidpunct}
{\mcitedefaultendpunct}{\mcitedefaultseppunct}\relax
\EndOfBibitem
\bibitem[Devereux \latin{et~al.}(2020)Devereux, Pezzella, Raghunathan, and
  Meuwly]{MM.dcm:2020}
Devereux,~M.; Pezzella,~M.; Raghunathan,~S.; Meuwly,~M. Polarizable Multipolar
  Molecular Dynamics Using Distributed Point Charges. \emph{J. Chem. Theo.
  Comp.} \textbf{2020}, \emph{16}, 7267--7280\relax
\mciteBstWouldAddEndPuncttrue
\mciteSetBstMidEndSepPunct{\mcitedefaultmidpunct}
{\mcitedefaultendpunct}{\mcitedefaultseppunct}\relax
\EndOfBibitem
\bibitem[Lee and Meuwly(2014)Lee, and Meuwly]{MM.oxy:2014}
Lee,~M.~W.; Meuwly,~M. Diffusion of atomic oxygen relevant to water formation
  in amorphous interstellar ices. \emph{Faraday Discuss.} \textbf{2014},
  \emph{168}, 205--222\relax
\mciteBstWouldAddEndPuncttrue
\mciteSetBstMidEndSepPunct{\mcitedefaultmidpunct}
{\mcitedefaultendpunct}{\mcitedefaultseppunct}\relax
\EndOfBibitem
\bibitem[Pezzella \latin{et~al.}(2018)Pezzella, Unke, and Meuwly]{MM.oxy:2018}
Pezzella,~M.; Unke,~O.~T.; Meuwly,~M. Molecular Oxygen Formation in
  Interstellar Ices Does Not Require Tunneling. \emph{J. Phys. Chem. Lett.}
  \textbf{2018}, \emph{9}, 1822--1826\relax
\mciteBstWouldAddEndPuncttrue
\mciteSetBstMidEndSepPunct{\mcitedefaultmidpunct}
{\mcitedefaultendpunct}{\mcitedefaultseppunct}\relax
\EndOfBibitem
\bibitem[Arasa \latin{et~al.}(2013)Arasa, van Hemert, van Dishoeck, and
  Kroes]{kroes:2013}
Arasa,~C.; van Hemert,~M.~C.; van Dishoeck,~E.~F.; Kroes,~G.-J. Molecular
  Dynamics Simulations of CO$_2$ Formation in Interstellar Ices. \emph{J. Phys.
  Chem. A} \textbf{2013}, \emph{117}, 7064--7074\relax
\mciteBstWouldAddEndPuncttrue
\mciteSetBstMidEndSepPunct{\mcitedefaultmidpunct}
{\mcitedefaultendpunct}{\mcitedefaultseppunct}\relax
\EndOfBibitem
\end{mcitethebibliography}

\clearpage

\begin{figure*}
\centering \includegraphics[width=0.8\textwidth]{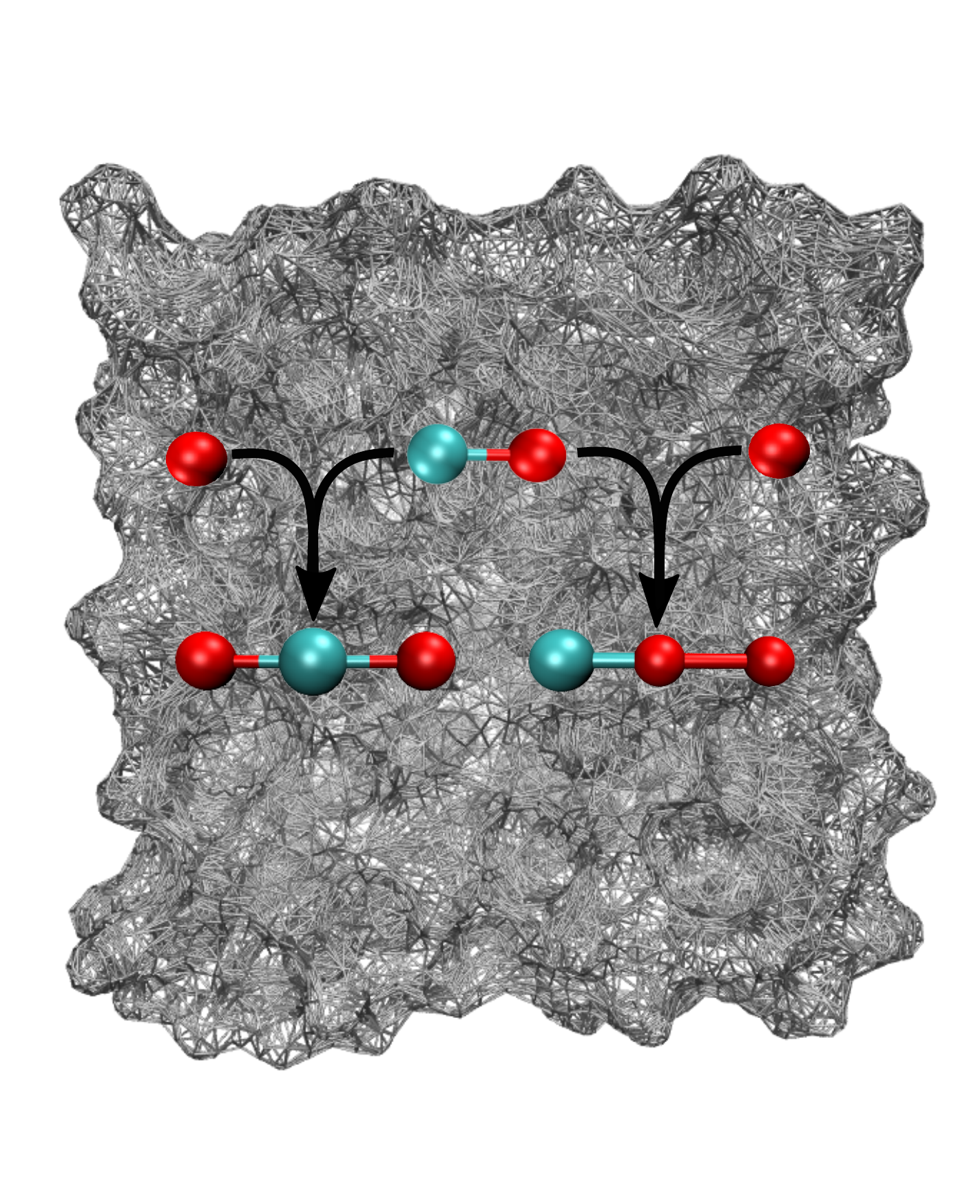}
\captionsetup{labelformat=empty}
\caption{TOC Graphic}
\end{figure*}

\end{document}


\section{Methods}
\subsection{The Energy Functions}
Reactive molecular dynamics simulations require potential energy
surfaces (PESs) that allow bond formation and bond
breaking.\cite{msarmd} Two such PESs were considered in the present
work. One was based on fitting Morse potentials $V(r) =
D_e(1-\exp{\left[-\beta(r-r_0)^2\right]})$ to the C--O interaction
based on {\it ab initio} calculations using the
MOLPRO\cite{molpro:2020} suite of programs at the MRCI/aug-cc-pVTZ
\cite{werner1988efficient, dunning1989gaussian} level of theory. The
fitting parameters are $r_0=1.164$ \AA\/, $D_0=173.50$ kcal/mol, and
$\beta=2.53$ \AA\/$^{-1}$. For the angular potential (the OCO bend)
the parameters were those from the CHARMM36 force field
\cite{best2012optimization} and this PES is referred to as MMH (for
Morse-Morse-Harmonic). Such an approach is similar to that used
previously for oxygen-oxygen recombination on amorphous solid
water.\cite{MM.oxy:2019,MM.o2:2020}\\

\begin{figure}[H]
\centering \includegraphics[scale=1.3]{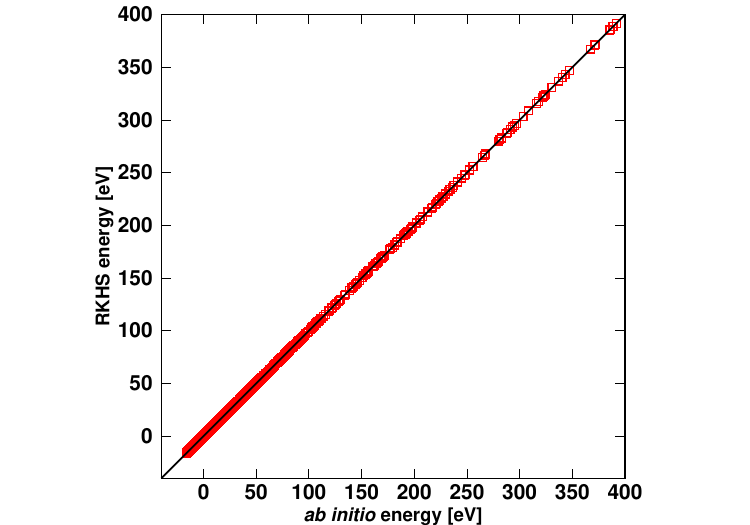}
\caption{Quality of the RKHS representation for the CCSD(T)-F12 points
  from the electron structure calculations. The RMSE is 0.0026 eV
  (0.06 kcal/mol) and the $R^2 = 1.0$.}
\label{sifig:pes}
\end{figure}

\noindent
The second ground state PESs is based on high-level
CCSD(T)-F12\cite{adler2007simple} calculations with the
aug-cc-pVTZ-F12\cite{peterson2008systematically} basis set using
MOLPRO.\cite{molpro:2020} For the geometries, a grid of $r$, $R$, and
$\theta$ values including $R \in [0.9, 5.9]$\AA\/, $r \in [0.8,
  2.1]$\AA\/, and $\theta \in [0, 180]^\circ$ was used. In total, 7800
reference energies were determined and represented as a reproducing
kernel Hilbert space.\cite{MM.rkhs:2017}. The root mean squared
difference between the reference energies and the RKHS representation
is 0.06 kcal/mol with $R^2 = 1.00$, see Figure \ref{sifig:pes}.\\

\noindent
For water the TIP3P model\cite{jorgensen-tip3p} was used and for
CO$_2$ the partial charges were $q_{\rm O} = -0.3e$ and $q_{\rm C} =
0.6e$ with standard van der Waals parameters from CHARMM. These
charges are consistent with those obtained from B3LYP/6-31G(d,p)
calculations snapshots from the MD simulations with CO$_2$ adsorbed to
the 10 nearest H$_2$O molecules which yield $q_{\rm C} = 0.73e$ and
$q_{\rm O} = -0.35e$. Simulations starting from the reactant state
(CO+O) were initialized with the same partial charges. This compares
with charges of $q_{\rm C} = 0.22e$ and $q_{\rm O} = -0.21e$ for the
CO molecule and $q_{\rm O} = -0.1e$ for an oxygen atom adsorbed to the
10 nearest H$_2$O molecules. To assess the dependence of the results
on the partial charges used, additional reactive MD simulations were
carried out with $q_{\rm O} = -0.1e$ and $q_{\rm C} = 0.2e$
(i.e. $q_{\rm CO} = 0.1e$) and with $q_{\rm O} = -0.2e$ and $q_{\rm C}
= 0.4e$ (i.e. $q_{\rm CO} = 0.2e$). In all cases, recombination was
found to speed up considerably compared with $q_{\rm CO} = 0.3e$ and
$q_{\rm O} = -0.3e$ due to the higher mobility of the CO molecule and
the O atom on the ASW when reduced partial charges are used.\\

\noindent
Although there are considerably more sophisticated models for
water,\cite{wang2013systematic,MM.dcm:2020} simulations with the TIP3P
model and point charges for atomic and molecular oxygen have yielded
satisfactory agreement for diffusion coefficients for atomic oxygen on
ASW with
experiments.\cite{MM.oxy:2014,MM.oxy:2018,MM.oxy:2019,MM.o2:2020}
Similarly, adsorption of CO to the ASW as reported\cite{kroes:2013}
from simulations with a quadrupolar model for CO together with TIP4P
is captured with the present parametrization. For temperatures up to
$T = 100$ K, a) CO with charges $q_{\rm C} = 0.22e$ and $q_{\rm O} =
-0.21e$ and b) $q_{\rm CO} = 0.3e$ and $q_{\rm O} = -0.3e$ remain
physisorbed to the ASW surface on the nanosecond time scale. Also, the
more elaborate parametrizations, in particular for water, are
computationally considerably more expensive and do not allow routine,
multiple $> 10$ ns simulations to be carried out. As the main focus of
the present work is on the CO+O recombination reaction the nonbonded
models used here were deemed sufficient.\\

\subsection{Molecular Dynamics Simulations}
\begin{figure}[H]
\centering
\includegraphics[scale=0.6]{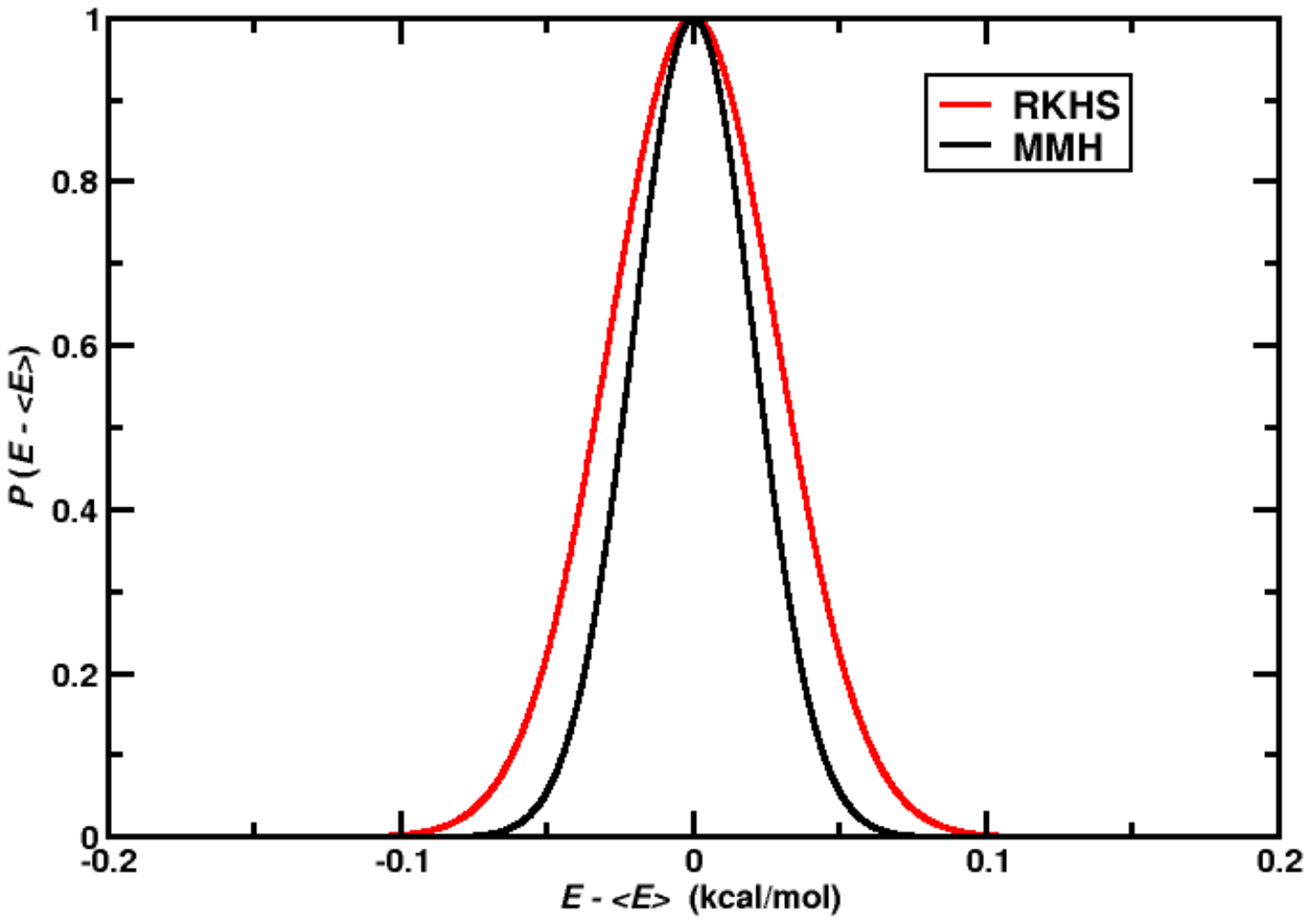}
\caption{Energy conservation for a reactive simulation using MMH
  (black) and the RKHS representation of the CCSD(T)-F12 calculations
  (red). The Gaussian shape of the curve confirms the energy
  conservation.}
\label{sifig:econs}
\end{figure}

In the following, the coordinates are the CO stretch $r$, the
separation $R$ between the center of mass of CO$_{\rm A}$ and O$_{\rm
  B}$ and $\theta$ is the O$_{\rm A}$CO$_{\rm B}$ angle. Hence $\theta
= 0$ corresponds to the COO structure whereas $\theta = 180^\circ$ is
the OCO conformation. Initial conditions were generated for a grid of
angles $\theta \in [67.5, 78.75, 84.375, 90.0, 101.25, 112.5, 123.75,
  135.0, 157.5, 180.0]^\circ$ and separations $R \in [2.66, 3.16,
  3.66, 3.91, 4.16, 4.66, 5.16, 5.66]$ \AA\/. A total of 80 initial
simulations were carried out to obtain initial coordinates and
velocities for each of the grid points. With constrained CO and O
position first 750 steps of steepest descent and 100 steps Adopted
Basis Newton-Raphson minimization was carried out, followed by 50 ps
heating dynamics to 50 K. Then, 100 ps equilibration dynamics was
carried out for 100 ps. From each of the runs coordinates and
velocities were saved regularly to obtain 1000 initial conditions for
each combination of angle and distance, i.e. in total 80000 initial
conditions. Production simulations 250 ps in length were then run from
saved coordinates and velocities with a time step of $\Delta t = 0.2$
fs in the $NVE$ ensemble. Energy conservation demonstrating the
correct implementation of the energies and forces for the reactive
PESs is reported in Figure \ref{sifig:econs} for both, simulations
with MMH and RKHS. \\

\newpage
\begin{sidewaystable}[!ht]
\caption{List of observations from the simulations with the MMH
  potential for combination of angles and distances where $R$ is the
  distance between CoM(CO) and O and $\theta$ is the angle between
  them.}
\begin{center}

  \begin{tabular}{c|c|c|c|c|c|c|c|c}
$\theta(^{\circ})$& $R$(\AA\/) & Total &  CO$_2$ formation & Atom exchange &  No collision complex & ES1 & ES2 & Both desorb\\
     67.5 & 2.66 & 100 & 0 & 0 & 1 &2 &3 &94\\
    \hline
     78.75& 2.66 & 500 & 6 & 20,17,12,6 & 107 & 96 & 23 &213\\
     78.75& 3.16 & 500 & 0 & 22,18,6,17 & 9 &8 &14 &406\\
     78.75& 3.66 & 100 & 0 & 1,3,3,2 & 2 &8 &0 &81\\
     \hline
     84.375& 2.66 & 500 & 36 & 89,25,21,45 & 47 & 37 &44 &156\\
     84.375& 3.16 & 100 & 1 & 2,6,2,2 & 7 & 2 & 6 & 72\\
     84.375& 3.66 & 100 & 0 & 6,7,4,5 & 6 & 6&1 &65\\
     \hline
     90.0 & 2.66 & 1000 & 179 & 73,40,32,64 &44 & 60&48 &460\\
     90.0 & 3.16 & 100 & 2 &10,8,2,4 &8 & 3&1 &62\\  
     90.0 & 3.66 & 100 & 0 &4,6,2,4 &8 & 7&0 &69\\
     90.0 & 3.91 & 100 & 0 &7,6,4,2 & 22 & 6&2 &51\\
     90.0 & 5.66 & 500 & 0 & 0 &0 &500 &0 &0\\
     \hline
      101.25 & 2.66 & 1000 & 751 & 44,37,9,4 & 0 &21 &5 &129\\
      101.25 & 3.16 & 100 & 8 & 11,5,2,9 & 20&6 &3 &36\\
      101.25 & 3.66 & 500 & 14 & 41,27,24,16 & 66 &52 &19 &241\\
      101.25 & 3.91 & 300 & 7 & 23,13,10,3 & 97 &33 &8 &106\\
      101.25 & 4.16 & 100 & 0 & 1,0,1,1 & 88 & 3 & 2 &4\\
      101.25 & 4.66 & 100 & 0 & 0  &99 &0 &0 &1\\
      \hline
      112.5 & 2.66 & 1000 & 913 & 19,2,1,0 & 46 &9 &4 &6\\
     112.5 & 3.16 & 500 & 205  & 76,20,10,12 &73 & 30 &12 &62\\
     112.5 & 3.66 & 200 & 42 & 34,11,4,8 & 18 &15 & 6 &62\\
     112.5 & 3.91 & 200 & 33 & 24,10,6,4 &44 &16 & 8&55\\     
     112.5 & 4.16 & 500 & 11 & 5,4,3,1 &441 &1 & 11&23\\
     \hline
     123.75 & 2.66 & 1000 & 998 & 0,1,0,0 & 0&0 &0 &1\\
     123.75 & 3.16 & 100 & 79 & 9,0,0,0 & 7 &2 &1 &2\\
     123.75 & 3.66 & 100 & 44 & 16,3,3,3 & 7 &5 &3 &16\\
     123.75 & 3.91 & 100 & 32 & 17,3,2,1 & 14 &3 &11 &17\\
     123.75 & 4.16 & 100 & 8 & 4,1,0,1 & 84 &0 &0 &2\\
     123.75 & 4.66 & 100 & 0 & 0 & 100 &0 &0 &0\\
     \hline
     135.0 & 2.66 & 1000 & 1000 & 0 & 0 & 0 & 0 & 0 \\
     135.0 & 3.16 & 1000 & 789 & 169,4,5,8 & 15 & 6&2 &2\\
     135.0 & 3.66 & 500 & 349 & 97,3,5,5 & 20 & 11 &3 & 7\\
     135.0 & 3.91 & 500 & 339  & 73,1,1,1 & 22 & 4 & 46 & 13\\
     135.0 & 4.16 & 1000 & 167  &14,1,2,6  & 253  & 35 &378 &144 \\
     135.0 & 4.66 & 1000 & 69 & 0 & 901 &0 & 27 &3  \\
     135.0 & 5.16 & 1000 &  0& 0 & 1000 &0 & 0 &0  \\
     135.0 & 5.66 & 1000 &  0& 0 & 1000 &0 & 0 &0       
       
     \end{tabular}
     \label{sitab:mmh1} 
   \end{center}
 \end{sidewaystable}

\newpage
\begin{sidewaystable}[!ht]
\caption{Continuation of Table \ref{sitab:mmh1}.}
\begin{center}
\begin{tabular}{c|c|c|c|c|c|c|c|c}
$\theta(^{\circ})$& $R$(\AA\/) & Total &  CO$_2$ formation & Atom exchange &  No collision complex & ES1 & ES2 & Both desorb\\
    \hline
    157.5 & 2.66 &   1000 & 1000 &0  &0  &0  &0  &0  \\
    157.5 & 3.16 &   500 & 500 &0  &0  &0  &0  &0  \\
     157.5 & 3.66 & 1000& 997&0  & 1 & 0& 1 & 1 \\
     157.5 & 3.91 &   500 & 494 &0  &3  &0  &2  &1  \\
     157.5 & 4.16 &   500 & 242 &0  &233  &0  &24  &1  \\
     157.5 & 4.66 & 500 & 0 &0  &500  &0  &0  &0  \\
     157.5 & 5.66 & 100& 0 & 0 &100  &0  &0  &0  \\
     \hline
     180.0 & 2.66 & 1000 & 1000 & 0  & 0 &0 & 0&0\\
     180.0 & 3.16 & 1000 & 1000 & 0  & 0 & 0& 0&0\\
     180.0 & 3.66 & 1000 & 1000  & 0  & 0 & 0 & 0&0\\
     180.0 & 3.91 & 1000 &991   &0   &9  &0 &0 &0\\
    180.0 & 4.16 & 500 &491  &0   &5  &3 &0 &1\\
     180.0 & 4.66 & 100 & 0  &0   &100  &0 &0 &0\\
     180.0 & 5.66 & 1000 & 0  & 0  & 1000 & 0& 0&0
    
\end{tabular}

In the atom exchange column, labels (a, b, c, and d) refer to
trajectories with atom exchange followed by a: CO and O remain on the
surface, b: CO and O desorb, c: O remains on surface and CO desorbs,
and d: CO remains on surface and O desorbs.
\label{sitab:mmh2} 
\end{center}
 \end{sidewaystable}

\newpage
\begin{sidewaystable}[!ht]
\caption{List of observations from the simulations using the 3d RKHS
  PES where $R$ is the distance between CoM(CO) and O and $\theta$ is
  the angle between them.}
\begin{center}
\begin{tabular}{c|c|c|c|c|c|c|c|c|c}
    
$\theta(^{\circ})$& $R$(\AA\/) & Total &  CO$_2$ formation & Atom exchange &  No collision complex & ES1 & ES2 & Both desorb & COO formation\\
\hline
 
   67.5 & 2.66 & 100 & 17 & 0 & 0 &0 &1 & 82 & 0\\
    \hline
    78.75& 2.66 & 100 & 51 & 0 & 3 & 1&0 &45 & 0\\
     78.75& 3.16 & 100 & 4 & 0 & 0 &0 &0 &2 & 94\\
     78.75& 3.66 & 100 & 0 & 0 & 89 &1 &0 &0 & 10\\
    \hline
     84.375& 2.66 & 100 & 62 & 0 & 2 & 5&3 &27 & 1\\
     84.375& 3.16 & 100 & 6 & 0 & 0 &0 &0 &0 & 94\\
     84.375& 3.66 & 100 & 0 & 0 & 96 &2 &0 &0 & 2\\
     \hline
     90.0 & 2.66 & 100 & 97 & 0 &3 &0 &0 &0 & 0\\
      90.0 & 3.16 & 100 & 45 & 0 &5 &0 &0 &0 & 50\\
     90.0 & 3.66 & 100 & 0 & 0 &0 &97 & 0& 0 &3\\
     \hline
      101.25 & 2.66 & 100 & 100 & 0 & 0 & 0 &0 & 0 &0\\
      101.25 & 3.16 & 100 & 97 & 0 & 3 & 0 &0 &0 &0\\
      101.25 & 3.66 & 100 & 0 & 0 & 96 &3 &0 &0 &1 \\
     \hline
     112.5& 2.66 & 100 &100 & 0 &0 &0 &0 &0 &0\\
     112.5& 3.16 & 100 & 100 & 0 & 0&0 &0 &0& 0\\
     112.5& 3.66 & 100 & 4 & 0 & 93&0 &1 &2& 0\\
     112.5& 3.91 & 100 & 0 & 0 & 94 &4 &1 &0 &1\\
     \hline
     123.75& 2.66 & 100 &100 & 0 &0 &0 &0 &0& 0\\
     123.75& 3.16 & 100 & 100 & 0 & 0&0 &0 &0& 0\\
     123.75& 3.66 & 100 & 5 & 0 & 93 &0 &0 &2&0\\
     123.75& 3.91 & 100 & 0 & 0 & 93 & 4 & 1 &0 &2\\
     \hline
     135.0 & 2.66 & 100 & 100  & 0 &0  &0  &0  &0  &0\\
     135.0 & 3.16 & 100 & 100  & 0 &0  &0  &0  &0  &0\\
     135.0 & 3.66 &500 &  46& 0 & 393 & 0&56 &0 & 5\\
     135.0 & 3.91 & 1000 & 1 &0  &731  &9 &242 &0 &17\\
     135.0 & 4.16 & 100 &  0& 0 &97  &0 &0 &0&3 \\
     
     \hline
     157.5 & 2.66 & 100 & 100  &0  &0  &0  &0  &0 &0 \\
     157.5 & 3.16 &100 &100  &0  &0  &0  &0  &0  &0\\
     157.5 & 3.66 & 100&  12 & 0 & 59 &0  &28  &  0 &1\\
     157.5 & 3.91 &100 &  0 & 0 &76  &0  &19  &0  &5\\
     \hline
     180.0 & 2.66 & 1000 & 1000& 0  & 0 &0 & 0&0 & 0\\
     180.0 & 3.16 & 100 & 99 & 0  &0  &0 &0 &1 & 0\\
     180.0 & 3.66 & 500 &53   &  0 & 292 & 2 &135 &8 &10\\
    180.0 & 3.91 & 100 &1   &  0 & 99 & 0 &0 &0 & 0\\
    180.0 & 4.16 & 100 &0   &  0 & 93 & 0 &5 &1 &1
 
\label{sitab:rkhs}     
\end{tabular}

\end{center}
\end{sidewaystable}

\clearpage

\begin{figure}[H]
\centering \includegraphics[scale=0.62]{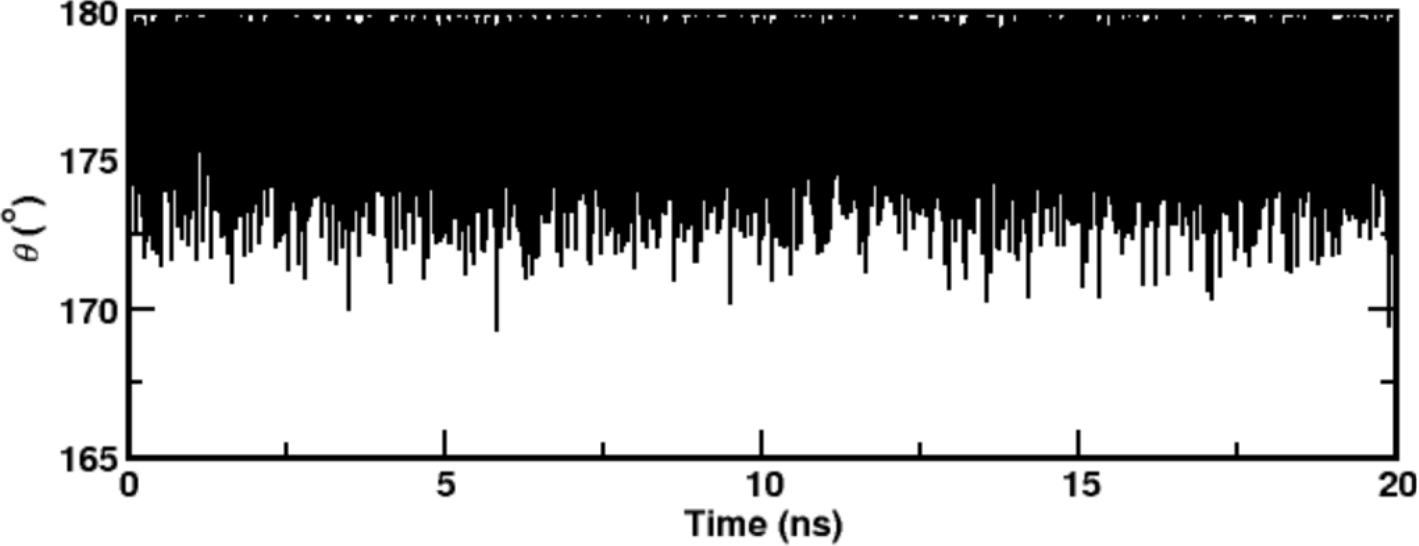}
\caption{Time series for $\theta(t)$ from 20 ns simulation, reporting
  the raw data of Figure 3B}.
\label{sifig:angle20ns}
\end{figure}

\begin{figure}[H]
\centering \includegraphics[scale=0.6]{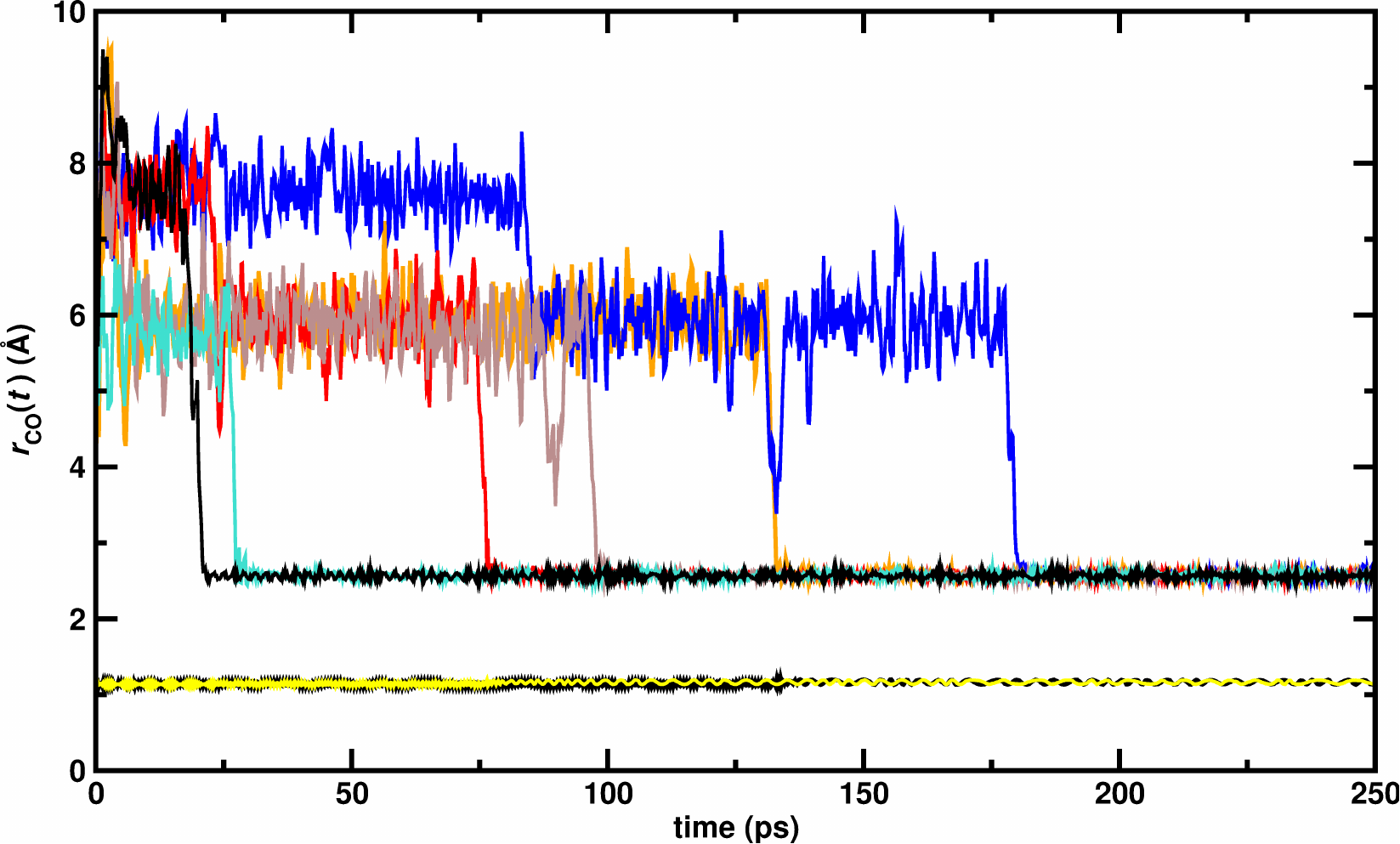}
\caption{Time series for rebinding into the COO conformation from
  simulations using the RKHS PES.}
\label{sifig:coo}
\end{figure}

\begin{figure}
\begin{center}
\includegraphics[scale=0.61]{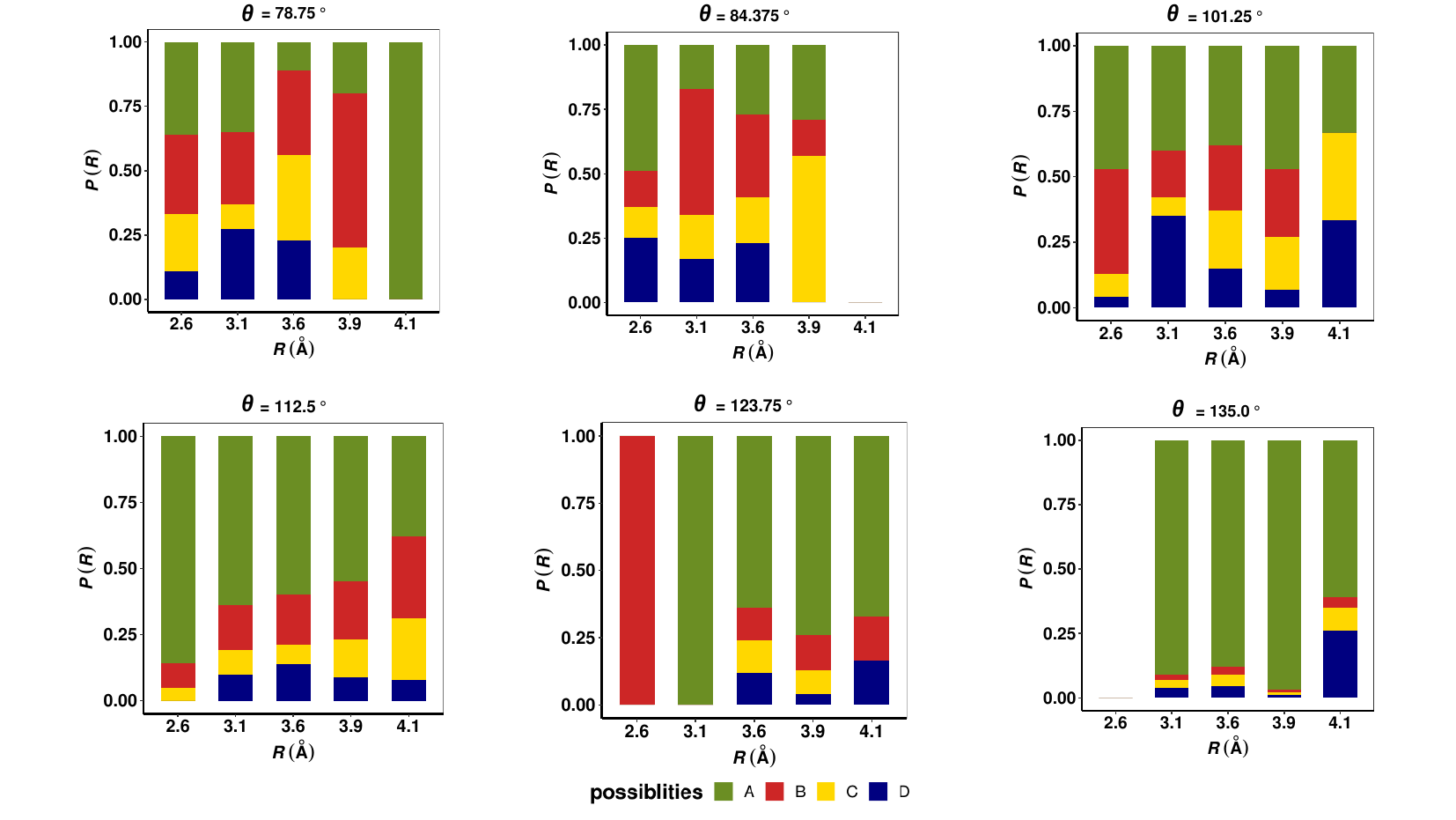}
\caption{After atom exchange, there are 4 further possibilities i.e
  both CO and O remains on the water surface (A), both CO and O
  desorbs (B), O remains and CO desorbs (C), CO remains and O desorbs
  (D) from water surface. Plots for atom exchange probability for each
  of the grid points using MMH.}
\label{sifig:channels}
\end{center}
\end{figure}

\clearpage

\bibliography{astro2}